\documentclass[preprint2]{emulateapj}
\usepackage{color}
\usepackage{amsmath}
\usepackage{threeparttable}
\usepackage{natbib}
\usepackage{subfigure}
\usepackage{times}

\DeclareGraphicsExtensions{.jpg,.pdf,.png,.eps,.ps}

\newcommand{\LCDM}{$\Lambda$CDM \,}
\newcommand{\LCDMx}{$\Lambda$CDM}

\newcommand{\nhat}{\hat{\mathbf{n}}}
\newcommand{\lvec}{\mathbf{l}}
\newcommand{\Lvec}{\mathbf{L}}
\newcommand{\dhat}{\hat{d}}
\newcommand{\clphi}{C_L^{\phi\phi}}

\newcommand{\dtwol}[1]{\frac{d^2 \lvec_{#1}}{(2\pi)^2}}
\newcommand{\dtwolprime}{\frac{d^2 \lvec'}{(2\pi)^2}}

\newcommand{\al}{A_\Lvec}
\def\muK{\mu{\mbox{K}}}
\newcommand{\lsplit}{\ensuremath{\ell_{\rm split}}}

\newcommand{\nlzero}{\ensuremath{N_L^{(0)}}}
\newcommand{\nlone}{\ensuremath{N_L^{(1)}}}
\newcommand{\nltwo}{\ensuremath{N_L^{(2)}}}

\newcommand{\hunit}{km\,s$^{-1}$\,Mpc$^{-1}$}

\newcommand{\omc}{\ensuremath{\Omega_ch^2}}
\newcommand{\sate}{\ensuremath{\sigma_8}}
\newcommand{\sumnu}{\ensuremath{\Sigma m_\nu}}

\newcommand{\omegac}{\Omega_ch^2}

\def \five {\textsc{ra5h30dec-55}}
\def \twthree {\textsc{ra23h30dec-55}}
\def \twonesix {\textsc{ra21hdec-60}}

\def \three {\textsc{ra3h30dec-60}}

\newcommand{\lsplittext}{$\ell$-split\ }
\def\fourfigshspace{-3.4}
\def\fourfigsscale{0.41}
\def\ell{l}
\newcommand{\chicmb}{\chi_{\textrm{\tiny CMB}}}
\newcommand{\alens}{\ensuremath{A_{\rm lens}}}
\newcommand{\alenszero}{\ensuremath{{A}_{\rm lens}^{0}}}
\newcommand{\comment}[1]{{}}

\def\McGill{1}
\def\KICPChicago{2}
\def\PhysicsUChicago{3}
\def\BCCP{4}
\def\UChicago{5}
\def\EFIChicago{6}
\def\AAUChicago{7}
\def\Argonne{8}
\def\Berkeley{9}
\def\NIST{10}
\def\Colorado{11}
\def\NASA{12}
\def\Davis{13}
\def\LBNL{14}
\def\Michigan{15}
\def\Munich{16}
\def\ExcellenceCluster{17}
\def\MPE{18}
\def\CaseWestern{19}
\def\Caltech{20}
\def\Minnesota{21}
\def\ArtInstChicago{22}
\def\Yale{23}
\def\CfA{24}

\begin{document}

\title{A measurement of gravitational lensing of the microwave background using South Pole Telescope data}

\author{
 A.~van~Engelen,\altaffilmark{\McGill}
 R.~Keisler,\altaffilmark{\KICPChicago,\PhysicsUChicago}
 O.~Zahn,\altaffilmark{\BCCP} 
 K.~A.~Aird,\altaffilmark{\UChicago}
 B.~A.~Benson,\altaffilmark{\KICPChicago,\EFIChicago}
 L.~E.~Bleem,\altaffilmark{\KICPChicago,\PhysicsUChicago}
 J.~E.~Carlstrom,\altaffilmark{\KICPChicago,\PhysicsUChicago,\EFIChicago,\AAUChicago,\Argonne}
 C.~L.~Chang,\altaffilmark{\KICPChicago,\EFIChicago,\Argonne}
 H.~M. Cho,\altaffilmark{\NIST} 
 T.~M.~Crawford,\altaffilmark{\KICPChicago,\AAUChicago}
 A.~T.~Crites,\altaffilmark{\KICPChicago,\AAUChicago}
 T.~de~Haan,\altaffilmark{\McGill}
 M.~A.~Dobbs,\altaffilmark{\McGill}
 J.~Dudley,\altaffilmark{\McGill}
 E.~M.~George,\altaffilmark{\Berkeley}
 N.~W.~Halverson,\altaffilmark{\Colorado}
 G.~P.~Holder,\altaffilmark{\McGill}
 W.~L.~Holzapfel,\altaffilmark{\Berkeley}
 S.~Hoover,\altaffilmark{\KICPChicago,\PhysicsUChicago}
 Z.~Hou,\altaffilmark{\Davis}
 J.~D.~Hrubes,\altaffilmark{\UChicago}
 M.~Joy,\altaffilmark{\NASA}
 L.~Knox,\altaffilmark{\Davis}
 A.~T.~Lee,\altaffilmark{\Berkeley,\LBNL}
 E.~M.~Leitch,\altaffilmark{\KICPChicago,\AAUChicago}
 M.~Lueker,\altaffilmark{\Caltech}
 D.~Luong-Van,\altaffilmark{\UChicago}
 J.~J.~McMahon,\altaffilmark{\Michigan}
 J.~Mehl,\altaffilmark{\KICPChicago}
 S.~S.~Meyer,\altaffilmark{\KICPChicago,\EFIChicago,\PhysicsUChicago,\AAUChicago}
 M.~Millea,\altaffilmark{\Davis}
 J.~J.~Mohr,\altaffilmark{\Munich, \ExcellenceCluster, \MPE}
 T.~E.~Montroy,\altaffilmark{\CaseWestern}
 T.~Natoli,\altaffilmark{\KICPChicago,\PhysicsUChicago}
 S.~Padin,\altaffilmark{\KICPChicago,\AAUChicago,\Caltech}
 T.~Plagge,\altaffilmark{\KICPChicago,\AAUChicago}
 C.~Pryke,\altaffilmark{\Minnesota }
 C.~L.~Reichardt,\altaffilmark{\Berkeley}
 J.~E.~Ruhl,\altaffilmark{\CaseWestern}
 J.~T.~Sayre,\altaffilmark{\CaseWestern}
 K.~K.~Schaffer,\altaffilmark{\KICPChicago,\EFIChicago,\ArtInstChicago}
 L.~Shaw,\altaffilmark{\Yale} 
 E.~Shirokoff,\altaffilmark{\Berkeley} 
 H.~G.~Spieler,\altaffilmark{\LBNL}
 Z.~Staniszewski,\altaffilmark{\CaseWestern}
 A.~A.~Stark,\altaffilmark{\CfA}
 K.~Story,\altaffilmark{\KICPChicago,\PhysicsUChicago}
 K.~Vanderlinde,\altaffilmark{\McGill}
 J.~D.~Vieira,\altaffilmark{\Caltech} and 
 R.~Williamson\altaffilmark{\KICPChicago,\AAUChicago} 
}

\altaffiltext{\McGill}{Department of Physics,
McGill University, 3600 Rue University, 
Montreal, Quebec H3A 2T8, Canada}

\altaffiltext{\KICPChicago}{Kavli Institute for Cosmological Physics,
University of Chicago, 5640 South Ellis Avenue, Chicago, IL, USA 60637}

\altaffiltext{\PhysicsUChicago}{Department of Physics,
University of Chicago,
5640 South Ellis Avenue, Chicago, IL, USA 60637}

\altaffiltext{\BCCP}{Berkeley Center for Cosmological Physics,
Department of Physics, University of California, and Lawrence Berkeley
National Labs, Berkeley, CA, USA 94720}

\altaffiltext{\UChicago}{University of Chicago,
5640 South Ellis Avenue, Chicago, IL, USA 60637}

\altaffiltext{\EFIChicago}{Enrico Fermi Institute,
University of Chicago,
5640 South Ellis Avenue, Chicago, IL, USA 60637}

\altaffiltext{\AAUChicago}{Department of Astronomy and Astrophysics,
University of Chicago,
5640 South Ellis Avenue, Chicago, IL, USA 60637}

\altaffiltext{\Argonne}{Argonne National Laboratory, 9700 S. Cass Avenue, Argonne, IL, USA 60439}

\altaffiltext{\Berkeley}{Department of Physics,
University of California, Berkeley, CA, USA 94720}

\altaffiltext{\NIST}{NIST Quantum Devices Group, 325 Broadway Mailcode 817.03, Boulder, CO, USA 80305}

\altaffiltext{\Colorado}{Department of Astrophysical and Planetary Sciences and Department of Physics,
University of Colorado,
Boulder, CO, USA 80309}

\altaffiltext{\NASA}{Department of Space Science, VP62,
NASA Marshall Space Flight Center,
Huntsville, AL, USA 35812}

\altaffiltext{\Davis}{Department of Physics, 
University of California, One Shields Avenue, Davis, CA, USA 95616}

\altaffiltext{\LBNL}{Physics Division,
Lawrence Berkeley National Laboratory,
Berkeley, CA, USA 94720}

\altaffiltext{\Michigan}{Department of Physics, University of Michigan, 450 Church Street, Ann  Arbor, MI, USA 48109}

\altaffiltext{\Munich}{Department of Physics,
Ludwig-Maximilians-Universit\"{a}t,
Scheinerstr.\ 1, 81679 M\"{u}nchen, Germany}

\altaffiltext{\ExcellenceCluster}{Excellence Cluster Universe,
Boltzmannstr.\ 2, 85748 Garching, Germany}

\altaffiltext{\MPE}{Max-Planck-Institut f\"{u}r extraterrestrische Physik,
Giessenbachstr.\ 85748 Garching, Germany}

\altaffiltext{\CaseWestern}{Physics Department, Center for Education and Research in Cosmology 
and Astrophysics, 
Case Western Reserve University,
Cleveland, OH, USA 44106}

\altaffiltext{\Caltech}{California Institute of Technology, MS 249-17, 1216 E. California Blvd., Pasadena, CA, USA 91125}

\altaffiltext{\Minnesota}{Department of Physics, University of Minnesota, 116 Church Street S.E. Minneapolis, MN, USA 55455}

\altaffiltext{\ArtInstChicago}{Liberal Arts Department, 
School of the Art Institute of Chicago, 
112 S Michigan Ave, Chicago, IL, USA 60603}

\altaffiltext{\Yale}{Department of Physics, Yale University, P.O. Box 208210, New Haven,
CT, USA 06520-8120}

\altaffiltext{\CfA}{Harvard-Smithsonian Center for Astrophysics,
60 Garden Street, Cambridge, MA, USA 02138}

\begin{abstract}
  We use South Pole Telescope data from 2008 and 2009 to detect the non-Gaussian
  signature in the cosmic microwave background (CMB) produced by
  gravitational lensing and to measure the power spectrum of the 
  projected gravitational potential.
  We constrain the
  ratio of the measured amplitude of the lensing signal to that
  expected in a fiducial \LCDMx\ cosmological model to be 
  $0.86 \pm 0.16$, with no lensing disfavored at $6.3 \sigma$.
  Marginalizing over \LCDMx\ cosmological models allowed by
  the Wilkinson Microwave Anisotropy Probe (WMAP7)
  results in a measurement of $\alens = 0.90 \pm 0.19 $, 
  indicating that the amplitude of matter fluctuations over the
  redshift range $0.5 \lesssim z \lesssim 5$ probed by CMB lensing is in good agreement
  with predictions.
  We present the results of several consistency checks.
  These include a clear detection of the lensing signature in CMB maps filtered 
  to have no overlap in
  Fourier space, as well as a ``curl'' diagnostic that is consistent
  with the signal expected for \LCDMx.  We perform a detailed study of
  bias in the measurement due to noise, foregrounds, and other effects and determine that these contributions are relatively small compared to the statistical uncertainty in the measurement.
  We combine this lensing measurement with results from WMAP7 to improve constraints    
  on cosmological parameters when compared to those from WMAP7 alone: we find a          
  factor of $3.9$ improvement in the measurement of the spatial curvature of the         
  Universe, $\Omega_k=-0.0014 \pm 0.0172$; a 10\% improvement in the amplitude of        
  matter fluctuations within \LCDMx, $\sigma_8=0.810 \pm 0.026$; and a 5\%               
  improvement in the dark energy equation of state, $w=-1.04\pm 0.40$.
  When compared with the measurement of $w$ provided by the combination of WMAP7 and external constraints on the Hubble parameter, the addition of the lensing data improve the measurement of $w$ by 15\% to give $w=-1.087\pm 0.096$.
\end{abstract}

\section{Introduction}

Measurements of the cosmic microwave background (CMB) have allowed us
to infer much about the Universe during the epoch of recombination
\citep[most recently,][]{jarosik10,das11b,keisler11}.  Over the last decade, measurements of interactions between the CMB and structures at lower redshifts have been used to constrain cosmology. 
These include the measurement of the effects of
reionization on the CMB at $z \sim 10$
\citep[e.g.,][]{kogut03a,komatsu11,zahn11b}; the detection of the
integrated Sachs-Wolfe effect from the onset of dark energy domination
at $z \sim 1$ \citep[e.g.,][]{fosalba03, padmanabhan05b}; and large
surveys using the Sunyaev-Zel'dovich (SZ) effect \citep{sunyaev72} 
to measure the growth of structure and the present day amplitude of  matter fluctuations  \citep[e.g.,][]{vanderlinde10, sehgal11, dunkley11,reichardt11,benson11}.

The gravitational lensing of the CMB is another long-promised source
of information on the post-recombination Universe
(e.g., \citealt*{blanchard87,cole89,seljak96b}; for a review see
\citealt*{lewis06}).  Lensing affects the CMB in two ways: it smooths
the CMB temperature power spectrum, and it correlates initially independent modes.  A
measurement of this latter effect can be obtained, for instance, with
the optimal quadratic estimator technique of \citet{seljak99}, \citet{hu01b}, and
\citet{okamoto03}, giving a reconstruction of the projected gravitational
potential \citep{bernardeau97, seljak99, zaldarriaga99, hu01b}.  Most
of the weight in the projection comes from high redshifts, with the maximum at $z \sim 2$.  CMB lensing measurements can
 thus probe the physics affecting structure formation
at high redshift, including the sum of the neutrino masses \citep{kaplinghat03, lesgourgues06}.  In addition, the
gravitational potential can be probed on very large scales, leading to
constraints on curvature, dark energy, and modified gravity models \citep{smith06b,calabrese09}.

Until recently, CMB observations had insufficient sensitivity and
angular resolution to detect lensing with the CMB alone.
\citet{smith07} instead performed a cross-correlation between a
reconstruction of the nearly full-sky CMB lensing field, obtained with
the third-year Wilkinson Microwave Anisotropy Probe (WMAP) data, and
the distribution of radio galaxies found in the NRAO VLA Sky Survey
(NVSS).  This resulted in a detection of the signature of CMB lensing
at $3.4\,\sigma$.  Similarly, \citet{hirata08} found a $2.5\,\sigma$
detection of cross-correlation between their lensing map, obtained
with the third-year WMAP release using a slightly different estimator,
and data from both the Sloan Digital Sky Survey and NVSS.

Experiments with smaller beam sizes and lower noise levels
have enabled  lensing detections using the CMB alone.
The Arcminute Cosmology Bolometer Array Receiver
\citep[ACBAR;][]{reichardt09a}, Atacama Cosmology Telescope
\citep[ACT;][]{dunkley11}, and South Pole Telescope
\citep[SPT;][]{keisler11} teams have found a preference for lensing in
the small-scale CMB temperature power spectrum at significances of $\sim
2\,\sigma$, $2.8\,\sigma$, and $5\,\sigma$, respectively.  Also,
although the WMAP satellite is not optimized for the study of the CMB power spectrum damping tail, which is most  affected by lensing, the
maps in the seventh-year WMAP release \citep{jarosik10} contain low enough noise that evidence for lensing using a novel kurtosis estimator was claimed at  2\,$\sigma$ by
\citet{smidt11}.  However, \citet{feng11} recently applied the optimal quadratic estimator of \citet{hu01b,okamoto03}
to the WMAP7 maps, finding no significant signal.
The first clear detection of the power spectrum of the CMB lensing potential was obtained by \citet{das11}, who used the quadratic estimator approach with ACT maps to obtain a $4\,\sigma$ detection.

Here, we perform quadratic lensing reconstruction and  present a detection of the lensing power spectrum from 590~square degrees of CMB sky observed by the SPT in 2008
and 2009.  This sky area is  approximately twice that used by \citet{das11}, and is observed with $\sim 25\%$ lower
noise. The paper is structured as follows.  In Section~2, we
review lensing of the CMB.  In Section~3, we briefly review the SPT
dataset, noting that we use the maps which were generated
for the CMB power spectrum analysis of \citet[][hereafter K11]{keisler11}. In Section~4, we detail our application of the
quadratic estimator technique to the SPT maps. In Section~5, we
estimate the impact of foregrounds and other systematic effects in the data,
and show that they can be controlled for the current level of precision.  In
Section~6, we present the quantitative results, including a measurement of the
amplitude of the lensing power spectrum relative to theoretical
expectations and  constraints on cosmological parameters. We conclude in Section~7.


\section{CMB Lensing}
\label{sec:theorybackground}
As CMB photons travel toward us, their paths are slightly deflected by 
fluctuations in the intervening matter density.  Since the
fluctuations are mostly in the linear regime on large scales, each
deflection is small, and we can consider the total deflection for a
given observation direction $\nhat$ as a sum of deflections along the
line of sight.  The projected potential for lensing $\phi(\nhat)$ is
then given by
\begin{equation}  \label{eq:redshiftintegral}
  \phi(\nhat) =  -2 \int_0^{\chicmb} d\chi \,   \frac{f_K(\chicmb - \chi)}{f_K(\chicmb)f_K(\chi)}
  \Phi(\chi \nhat, \chi),
\end{equation}
where $\Phi(\mathbf{r}, \eta)$ is the three-dimensional gravitational potential at position $\mathbf{ r}$ and conformal look-back time $\eta$, both measured with us at the origin, $\chi$
is the comoving distance along the line of sight, $\chicmb \simeq
14\,$Gpc is the comoving distance to the CMB, and $f_K(\chi)$ is the comoving angular diameter distance, with $f_K(\chi) = \chi$ in
a spatially flat universe.  Lensing shifts
 the unlensed CMB temperature $T^U(\nhat)$ at a sky
position $\nhat$ by the gradient of this lensing
potential, resulting in an observed CMB temperature 
\begin{align} 
  T(\nhat) =& T^U(\nhat + \nabla \phi(\nhat)) \nonumber\\ =&
  T^U(\nhat) + \nabla T^U(\nhat) \cdot \nabla \phi(\nhat) + \ldots.
\label{eq:lensingformula}
\end{align}
The statistics of the Gaussian, unlensed temperature field are
determined purely by the unlensed CMB power spectrum $C_\ell^U$ according
to
\begin{equation}  
\label{eq:diagcoupling}
\langle
T^U(\lvec_1) T^U(\lvec_2) \rangle = (2\pi)^2 \delta(\lvec_1 + \lvec_2)
C_{\ell_1}^U, 
\end{equation}
where we use the flat-sky Fourier convention
\begin{equation}
T(\lvec) = \int d^2 \nhat T(\nhat) e^{-i \lvec \cdot \nhat} ,
\end{equation}
and apply the high-$\ell$ limit in which the all-sky power spectrum
$C_\ell$ becomes equivalent to its flat-sky Fourier analogue.

When the CMB is lensed, the coupling 
between the CMB gradient and the $\phi$ gradient in
Eq.~\ref{eq:lensingformula} leads to an
off-diagonal correlation between CMB multipole moments of
\begin{align} \label{eq:lensmodecoupling}
\langle T(\lvec_1) T(\lvec_2) \rangle = & \Lvec \cdot (\lvec_1
C_{\ell_1}^U + \lvec_2 C_{\ell_2}^U) \phi(\Lvec) \nonumber \\
\equiv & f(\lvec_1, \lvec_2) \phi(\Lvec) ,
\end{align}
 at linear order in $\phi$.  Here, $\Lvec = \lvec_1 + \lvec_2 $, and we have assumed $\lvec_1 \ne
-\lvec_2$.\footnote{Throughout the paper, we use capital $\Lvec$ to refer to the argument of the lensing field, and lowercase $\lvec$ to refer to the argument of the CMB temperature field.}

A quadratic estimator takes advantage of this off-diagonal
coupling by averaging over products of pairs of observed CMB modes $T(\lvec_1)$
and $T(\lvec_2)$ that satisfy $\lvec_1 + \lvec_2 = \Lvec \ne 0$ to
reconstruct a mode $\phi(\Lvec)$.  This is distinct from 
lensing detections using the CMB temperature power spectrum
\citep{calabrese08, reichardt09a,  das11b, keisler11}, which probe the
effects of lensing on ``on-diagonal'' CMB modes with $\lvec_1 +
\lvec_2 = 0$. 

Although the typical deflection angle is small, $|\nabla \phi|_{\rm
  RMS} \simeq 2.4\arcmin$, the lensing deflection field is coherent across
several degrees on the sky.  The mode coupling is thus strongest for
pairs of modes in the CMB map with small vectorial separation, of magnitude $L \la
1000$.  Additionally, the signal is most significant on scales at
which the CMB temperature power spectrum is a steep function of $\ell$
(e.g., \citealt{zahn06,bucher10}). This condition is met on small scales (high $\ell$),
where the primary CMB fluctuations are exponentially damped due to
photon diffusion effects during recombination.  Thus, an ideal lensing
estimate will search for nonzero coupling between pairs of $\ell > 1000$ CMB
modes, separated by $\Lvec$ of several hundred.  
Data from the current generation of CMB temperature experiments, namely SPT, ACT, and the \emph{ Planck} Surveyor, can be used to resolve the fluctuations in the damping tail region of the CMB temperature power spectrum.  These data are thus well-suited for
measuring the lensing signal on degree scales, using only the measured fluctuations on scales of several arcminutes.  


\section{The South Pole Telescope}

The SPT is an off-axis Gregorian telescope with a 10-meter diameter
primary mirror located at the South Pole.  The receiver is equipped
with 960 horn-coupled spiderweb bolometers with superconducting
transition-edge sensors.  The detectors are divided between three
frequency bands centered at $95$, $150$, and $220$~GHz.  The telescope
and receiver are discussed in more detail in \citet{ruhl04},
\citet{padin08}, and \citet{carlstrom11}.

\subsection{Survey and Fields}
\label{ref:fields}
The SPT-SZ survey is a multi-year observation program with the principal goals of using the Sunyaev-Zel'dovich effect to produce a nearly mass-limited sample of galaxy clusters for cosmological studies, e.g., for measuring the growth of structure to constrain the dark energy equation of state \citep{staniszewski09, vanderlinde10,foley11, williamson11} and to measure the power spectrum of
the mm-wave sky on small angular scales
(\citealt{lueker10,hall10,shirokoff11}; K11; \citealt{reichardt11}).

In this work, we use only the  $150$~GHz due to the lower noise, of
approximately 18\,$\mu$K-arcmin\footnote{Throughout this work,
map signal and noise amplitudes are expressed in units of 
K-CMB, expressing deviations from the average measured intensity as
equivalent temperature fluctuations in the CMB.}.  We use the two fields observed by the
SPT in 2008, which total 197 square degrees, and two of the fields
observed in 2009, which total 393 square degrees.  These four fields
correspond to the fields marked \five, \twthree, \three, and
\twonesix\ in Table 1 of K11.

\subsection{Mapmaking from time streams}
\label{sec:mapmaking}

The SPT maps used in this analysis are identical to those used by K11.  The processing used to go from time-ordered data (TOD) to maps is described in more detail in that work, and we summarize the main points here.  First, the raw 100 Hz TOD are low-pass filtered at 7.5\,Hz and resampled at 16.7\,Hz.  Next, the TOD are band-pass filtered by applying a second low-pass filter at 5\,Hz and by removing a Legendre polynomial from each scan across the field.  Approximately 1.5 degrees of freedom are removed per degree on the sky.  Finally, we subtract the mean signal across each detector module\footnote{The SPT array consists of 6~wedge-shaped bolometer modules, each with 160~detectors.  Each wedge is configured with a set of filters that determine its observing frequency ($95$, $150$, or $220$~GHz).} at each time sample.  This spatial high-pass filter removes atmospheric noise that is correlated among detectors.

The filtered TOD, in conjunction with the pointing information, are
projected onto two-dimensional maps using the oblique Lambert
equal-area azimuthal projection \citep{snyder87} with pixels of size
$1\arcmin$.

All SPT temperature power spectrum analyses are performed using cross-power estimates between  maps of disjoint observations of each SPT field, with each observation consisting of several hours of time-ordered data.  By contrast, in the lensing analysis we use only the season-averaged map for each field.

\subsection{Source Removal}
\label{sec:sources}
To remove bright sources from the maps, we center a square mask
on each source and ``paint in'' CMB fluctuations using interpolation.  

We first derive a source list for masking extremely bright sources
and galaxy clusters using SPT catalogs (compiled using maps that were
processed slightly differently and with smaller pixels).  A $12\arcmin
\times 12\arcmin$ mask is applied to positive sources brighter than
40$\,\sigma$ and galaxy clusters brighter than 20$\,\sigma$. The source
densities of extremely bright sources and galaxy clusters are $\sim
0.2$ and $0.01$ per square degree, respectively.  We discuss the dependence on masking levels in Section~5.

Next, fainter sources are identified by applying a matched filter to the
maps used for the lensing analysis (which are not optimized for point
source detection) and selecting all sources above 6$\,\sigma$.  These
sources are removed by masking the surrounding $8\arcmin \times
8\arcmin$ region.  The effective flux cut is approximately 10\,mJy,
with a typical source density of 0.5~sources per square degree.

An estimate of the CMB fluctuations in the masked region is then
determined in a $16\arcmin$ square surrounding each source region,
where a Gaussian random field  is assumed.  The covariance matrix
${\bf C}$ between pixels in the $16\arcmin$ square is calculated from
a large number of $16\arcmin$ square regions of SPT maps.  A new
matrix ${\bf C'}$ is constructed by setting the diagonal elements
corresponding to the masked pixels to large values.  Finally, the
interpolated map is estimated as ${\rm T_{est}}= {\bf C C'}^{-1} {\rm T_{map}}$,
where ${\rm T_{map}}$ is the original map.
This procedure is a variant of Wiener filtering
\citep[e.g.,][]{knox98b}.  It is analogous to maximum likelihood
mapmaking for a small subregion that has high noise embedded in a
larger well-measured map that has known large-scale correlations.  It
is also similar to a constrained realization in the masked region
\citep[e.g.,][]{hoffman91}, but differs in that there is no random
noise added on small scales.

The fractional residual power from a point source after this masking
procedure is applied is less than $6 \times 10^{-4}$; the total fraction of sky area masked is  $\sim 1\%$. 
We perform the same procedure on the
simulated observations described below to include their impact on the
final results.

\subsection{Beam}

The beam shape is
measured using a combination of observations of planets and bright
point sources.  The central beam is approximately Gaussian and is
measured using the five brightest point sources in the fields observed
by SPT in 2008 and 2009.  This approach naturally takes into account the
effective beam enlargement due to random errors in the pointing
reconstruction.  The outer beam, which accounts for roughly 15\% of
the total beam solid angle, is measured with planet observations.  The
SPT beams are described in more detail in K11.
We discuss the impact of the beam uncertainties on the lensing results
in Section~\ref{sec:beamuncert}.

\section{Lensing Analysis}
\label{sec:analysis}

In this section, we describe our procedure for measuring the
lensing signal. We discuss our method for making lensing maps, 
give an overview of the detailed end-to-end simulations that
play a central role in the analysis, and describe the methods
for characterizing the lensing power spectrum.

\subsection{Estimating lensing maps from CMB maps}
\label{sec:qe}
Several methods have been
proposed to detect the lensing signature in CMB maps. 
 The most well-studied estimators reconstruct lensing using
the coupled nature of CMB modes in the Fourier domain
\citep{bernardeau97,zaldarriaga00b}, which can be optimized with
particular choices of filters \citep{hu01c, hu01b, hu02a, okamoto03}.
Pixel-space versions of these estimators have also been formulated
\citep{bucher10, carvalho10}.  Alternative approaches involve
formulating and maximizing a lensing likelihood function in pixel space \citep{
  hirata03a, hirata03b,anderes11}.  However, for temperature-only data
with current noise levels, these maximum-likelihood approaches are not
expected to lead to appreciable gains in signal-to-noise ratio.  Therefore, in
this work we  use the quadratic formulation to isolate the
signature of lensing in the four-point function in the Fourier domain.
The same approach was taken in previous detections \citep{smith07,
  hirata08, das11}.

As shown by \citet{hu01b}, the quadratic combination
that maximizes the lensing measurement signal-to-noise is the
temperature-weighted gradient, in which one multiplies a
filtered CMB gradient with a high-pass filtered CMB map.  The
filters are chosen to weight the observed CMB according to the inverse
variance, and to select for the mode-coupling in
Eq.~\ref{eq:lensmodecoupling}. The gradient-filtered map takes the form
\citep{hu01c, hu01b}
\begin{equation}
\label{eq:gradientfilter}
 \mathbf{G}(\nhat) = \int \dtwol{}  \frac{ C_\ell^U}{ C_\lvec^t} 
i\lvec T(\lvec) e^{i\lvec \cdot \nhat},
\end{equation}
and the high-pass-filtered map takes the form
\begin{equation}
\label{eq:lensfilter}
W(\nhat) = \int \dtwol{} \frac{1}{C_\lvec^t} T(\lvec) e^{i\lvec
  \cdot \nhat}.
\end{equation}
Here,  $C_\ell^U$ denotes the unlensed CMB power spectrum and $C_\lvec^t =
C_\ell^L + C_\lvec^N + C_\ell^F  $ denotes the total power in the
observed CMB map.  The components of this total power include the lensed CMB temperature power $C_\ell^L$, the
 noise power $C_\lvec^N$, and the power spectrum of the foregrounds $C_\ell^F$.  We denote the argument of the noise power with a vector, $\lvec$, due to its anisotropic nature as described below.
After forming the real-space product $\mathbf{G}(\nhat) W(\nhat)$, one takes the filtered divergence of the
result.  This leads to an estimate for the scalar lensing deflection field $d(\Lvec)$,
\begin{equation}
\dhat(\Lvec) = - {\al \over L} \int d^2\nhat \, \nabla \cdot (\mathbf{G}(\nhat)
W(\nhat)) e^{-i \Lvec \cdot \nhat},
\end{equation}
where $A_\Lvec$ is the normalization.  The deflection field is  related to the lensing potential through $d(\Lvec) = L\phi(\Lvec)$.  
Expressed in Fourier space, the estimator is
\begin{equation}
\dhat(\Lvec) =  \frac{A_\Lvec}{L} \int \dtwol{1} F(\lvec_1, \Lvec -
\lvec_1) T(\lvec_1) T(\Lvec - \lvec_1),
\end{equation}
where the filter function
\begin{equation}  \label{eq:filter}
F(\lvec_1, \lvec_2) = \frac{f(\lvec_1, \lvec_2) }{2 C_{\lvec_1}^t
  C_{\lvec_2}^t}.
\end{equation}
The factor $A_\Lvec$ is chosen to normalize the estimate such that to
linear order in $\phi$
\[
\langle \hat{d}(\Lvec)\rangle_\textrm{\small CMB} = L \phi(\Lvec), 
\]  
according to Eq.~\ref{eq:lensmodecoupling}.  
The subscript ``CMB'' on
the expectation value indicates that the ensemble average is taken
over a set of CMB realizations all lensed by the
same $\phi$ field.  The function $A_\Lvec$ is given, in the absence of data windowing, by
\begin{equation}     \label{eq:weighting}
 A_\Lvec = \left (\frac{1}{L^2} \int \dtwol{1} \frac{(\Lvec \cdot
   \lvec_1 C_{\ell_1}^U + \Lvec \cdot \lvec_2 C_{\ell_2}^U)^2}{2
   C_{\lvec_1}^t C_{\lvec_2}^t} \right )^{-1}.
\end{equation}
This function is also the noise in the lensing estimate \citep{hu01b,kesden03}.

In our simulation-based approach, detailed below, the results of the
lensing estimate are not sensitive to the exact settings for the CMB
power spectra in the filter; a mismatch between the assumed and exact power
will lead to a small loss in optimality, but no bias.  We obtain the
total CMB power in the denominator, $C_\lvec^t$, assuming
contributions from a lensed WMAP7 best-fit CMB power spectrum
\citep{komatsu11}; power from uncorrelated point sources of $C_\ell
\equiv 7 \times 10^{-6}\,\mu$K$^2$; a flat-bandpower component with
$\ell^2C_\ell/2\pi \equiv 10\,\mu$K$^2$ (designed to capture the
combination of the power spectra of the thermal SZ effect, the kinetic
SZ effect, and clustered dusty galaxies); and finally, a term due to
instrumental noise.  The two-dimensional noise power spectra of the
maps, $C_\lvec^N$, are calculated directly from ensembles of SPT
difference maps.  We obtain pseudo-independent SPT noise realizations,
containing little response to any on-sky signal, by flipping the signs
of half of the several hundred observations of each field and then
performing a coadd.

Due to the SPT observing strategy, in which the telescope scans in azimuth
between steps in elevation, the noise power in the maps, $C_\lvec^N$,
is anisotropic.  In particular, the noise is substantially larger at
low values of $\ell_x$, the Fourier conjugate to the scan direction\footnote{Due to the location of the observatory at the South Pole,  scans in the azimuthal direction are equivalent to scans in right ascension; there is no relative rotation between celestial and telescope co-ordinates on the sky.}.
The lensing estimation procedure, described above, naturally down-weights these modes.

For the given map noise levels of roughly $18\,\muK-{\rm arcmin}$, the
lensing signal is concentrated in the annular range of CMB temperature
multipoles $1200 \la \ell \la 3000$.  We apply an
azimuthally-symmetric bandpass filter to isolate these modes.  Isolating the temperature modes $T(\lvec)$ within this annulus is effectively equivalent to setting the denominator of the filter in Eq.~\ref{eq:filter}, $C_\lvec^t$, to be infinite outside this annulus.

The instrumental time stream filtering discussed in
Section~\ref{sec:mapmaking} leads to a signal transfer function which
must be carefully evaluated.   This is particularly true at low values
of $\ell_x$, which are more aggressively filtered.  Although this
transfer function was characterized down to low values of
$\ell_x$ in K11, with the lowest reported bin at $\ell = 650$, in this
work we do not consider CMB modes with $\ell_x < 1000$ for simplicity.
We estimate that a reduction of this cut to lower $\ell_x$ would
increase the total lensing detection significance by up to $\sim
10\%$.

Together, the anisotropic noise and $l_x$ filtering lead to an
anisotropic function $A_\Lvec$, with higher amplitude in the $L_y$-direction
than in the $L_x$-direction by a factor of $\sim 4$.

\subsubsection{Curl test}
\label{sec:curlestimator}
As described in the previous section, the lensing estimate is derived
by taking the divergence of a vector map, ${\bf G}(\hat{\bf n})
W(\hat{\bf n})$.  This estimate is constructed for a field which has a
gradient component, but no curl, as expected for the lensing deflection
field.  By taking the curl instead, we can construct an estimator that
closely resembles the lensing estimator, but is optimized for
curl-like sources \citep{cooray05}. The estimate is formulated as
\begin{equation}
\hat{c}(\Lvec) = - {A_\Lvec^c \over L} \int d^2\nhat \, \nabla \star (\mathbf{G}(\nhat)
W(\nhat) )e^{-i \Lvec \cdot \nhat}, 
\end{equation}
with the factor
\begin{equation}     \label{eq:curlweighting}
  A_\Lvec^c = \left ( \frac{1}{L^2} \int \dtwol{1} \frac{(\Lvec \star \lvec_1
    C_{\ell_1}^U + \Lvec \star \lvec_2 C_{\ell_2}^U)^2}{2 C_{\lvec_1}^t
    C_{\lvec_2}^t} \right )^{-1}.
\end{equation}
 The operator $\star$ is defined via $\mathbf{A} \star
\mathbf{B} = A_yB_x - A_xB_y$.

This quantity is analogous to estimating ``B-modes'' in cosmic shear
experiments. 
However, many foregrounds (e.g., point sources) have
negligible contribution to the curl, and as discussed in Appendix~\ref{sec:curlappendix}, gravitational lensing actually generates a
non-zero curl power spectrum when using a quadratic estimator.  The
curl estimate is most useful as a test of our understanding of the
fluctuation power in the maps.

\subsubsection{Apodization}
\label{sec:apodization}

\begin{figure*}[htb]
  \plotone{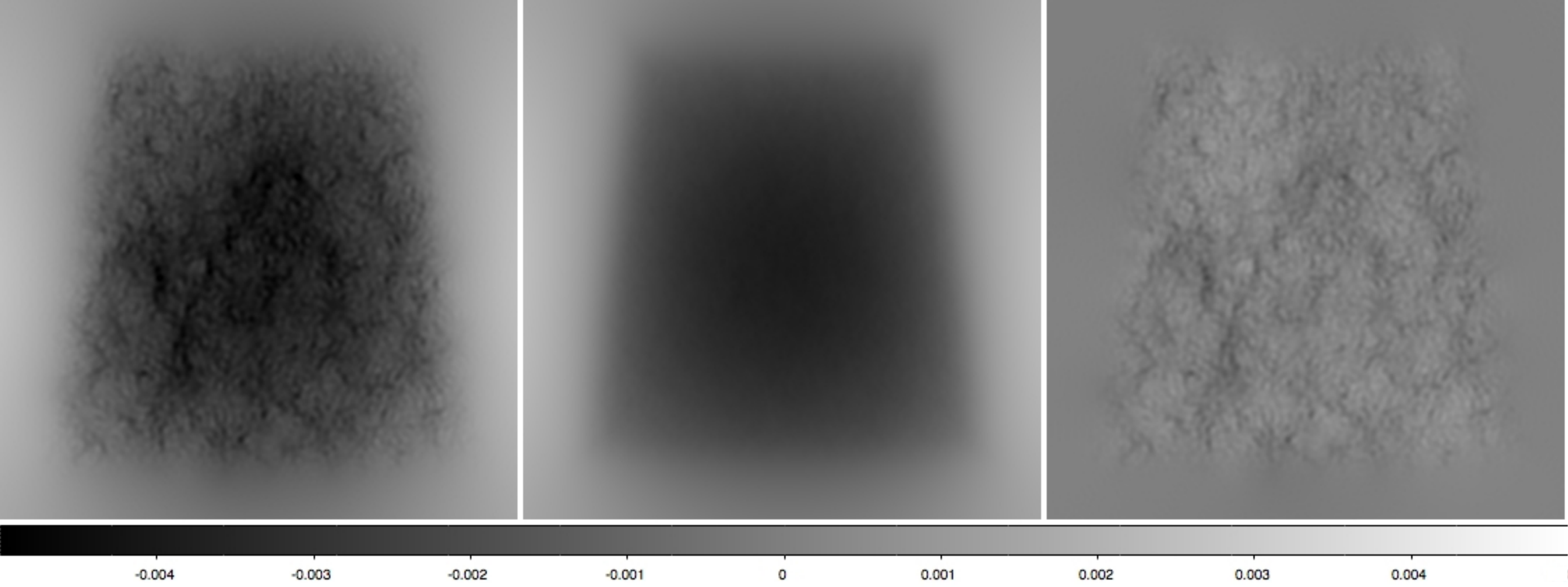}
  \caption{Impact of apodization: (left) reconstruction of lensing
    deflection for one of the SPT fields (\five); (middle) mean
    estimated deflection for 100 simulations, indicating the mean
    apodization feature; (right) resulting estimate of the deflection in  the SPT field after subtracting the estimated apodization feature. All maps have the same
    greyscale ($\pm$0.005). }
\label{fig:apodization_maps}
\end{figure*}

Previous work has dealt with correlations due to sky cuts by
setting the noise to be large in cut
pixels, and then taking the full pixel-pixel covariance matrix into
account \citep{smith07}.  
In a simpler, sub-optimal approach we formulate the estimate
initially neglecting the apodization, and then characterize the apodization 
response using Monte Carlo simulations.

Given that the observed CMB has been convolved with the Fourier
transform of an apodization window,   $R(\nhat)$, i.e.,\
\begin{equation}  
  T_R(\lvec) = \int \dtwolprime T(\lvec') R(\lvec - \lvec'),  
\end{equation} 
it will possess off-diagonal correlations given by 
\begin{equation}
  \langle T_R(\lvec_1) T_R(\lvec_2) \rangle = \int \dtwolprime
  C_{\lvec'}^t R(\lvec_1 - \lvec') R(\lvec_2 + \lvec'). 
\end{equation}
Running such a CMB field through the lensing estimator will result in a
spurious signal at $\Lvec \ne 0$ given by
\begin{align}  \langle \dhat(\Lvec) \rangle = & \frac{A_L}{L} \int
  \dtwol{1} \int \dtwolprime \nonumber  F(\lvec_1, \Lvec -
  \lvec_1) \nonumber \\ &\times C_{\lvec'}^t R(\lvec_1 - \lvec')
  R(\Lvec - \lvec_1 + \lvec').  \end{align} This signal is a weighted
average over the total power in the map, $C_\lvec^t$, and is present
even for an unlensed CMB.  In the limit of a very broad window in real
space, $R(\lvec) \rightarrow \delta(\lvec)$ and this
signal goes to zero for $\Lvec \ne 0$.  For a typical apodization
window whose size is tens of degrees, we find that the signal falls to
zero quickly with $L$.

In practice, we characterize and remove the mean apodization feature
using Monte Carlo simulations.  Figure~\ref{fig:apodization_maps}
shows the deflection map in real space, $d(\nhat)$, for one of the SPT
fields (\five) before and after subtraction of this feature.  The
feature takes on numerical values $\sim 5$ times a typical $d$
fluctuation. However, as it is a slowly-varying function, it is
largely decoupled from the multipoles $L > 100 $ at which we report
our lensing results.

\subsection{SPT Lensing Simulations}
\label{sec:lenssims}
To characterize the impact of filtering choices on the lensing signal,
we use end-to-end simulations of the data and lensing estimation.

We first generate 100 simulated full-sky lensed CMB realizations using
the LensPix package \citep{lewis05}, up to a maximum multipole of
$\ell_{\rm{max}} = 5000$ and at $0.8\arcmin$ resolution. The lensing field is taken to be Gaussian; we address the impact of non-Gaussianities in the lensing field, due to the effects of nonlinear growth of density fluctuations, in Section~\ref{sec:nonlinear}.    We also generate the same number of unlensed
full-sky simulations using the HEALPix tools \citep{gorski05}\footnote{http://healpix.jpl.nasa.gov}.  The
unlensed CMB simulations are constructed to have the same power spectrum as
the lensed simulations, but do not contain the lensing-induced mode-couplings.  The
cosmological parameters are given by the best-fit model for WMAP7
together with the high-multipole measurements of the ACBAR
\citep{reichardt09a} and QUaD \citep{brown09, friedman09}
experiments as found on the LAMBDA website\footnote{http://lambda.gsfc.nasa.gov/product/map/dr4/parameters.cfm}.  This model consists of physical baryon density $\Omega_bh^2 = 0.02235$, physical cold dark matter density $\Omega_ch^2 = 0.1086$, Hubble parameter $H_0 = 70.92\,$\hunit, optical depth to recombination $\tau = 0.0878$, amplitude of primordial scalar fluctuations $A_s = 2.453\times10^{-9}$ and spectral index of primordial scalar fluctuations $n_s = 0.960$.  The latter two quantities are quoted at reference wavenumber of $k=0.002\,$Mpc$^{-1}$.

We then project portions of these CMB simulations for each field onto
the flat sky, using the oblique equal-area Lambert projection as was
done for the real data.  To these CMB fields we then add discrete
point sources and Gaussian backgrounds consistent with those expected
from the Sunyaev-Zel'dovich effect from galaxy clusters and cosmic
infrared background fluctuations
\citep{hall10,dunkley11,reichardt11,planck11-6.6_arxiv}, as was done
in K11.  We then pass these simulated fields through simulated
observations which take into account the detailed time stream filtering
applied to the real data, as described in K11.  
Noise realizations are generated by differencing real SPT observations
in two ways: data taken when the telescope is scanning in different
directions in azimuth are assigned opposite signs, and then data from
separate observations are also assigned random signs. 

We then perform lensing reconstruction on these simulated maps, setting the CMB lensing filters to mimic the procedure applied to
the real SPT maps, including point source removal and apodization.

\subsection{Estimating lensing power spectra from lensing maps}
In this subsection, we describe our method of estimating the power spectra of the reconstructed deflection maps, including our approach for treating noise bias.

\begin{figure}
\plotone{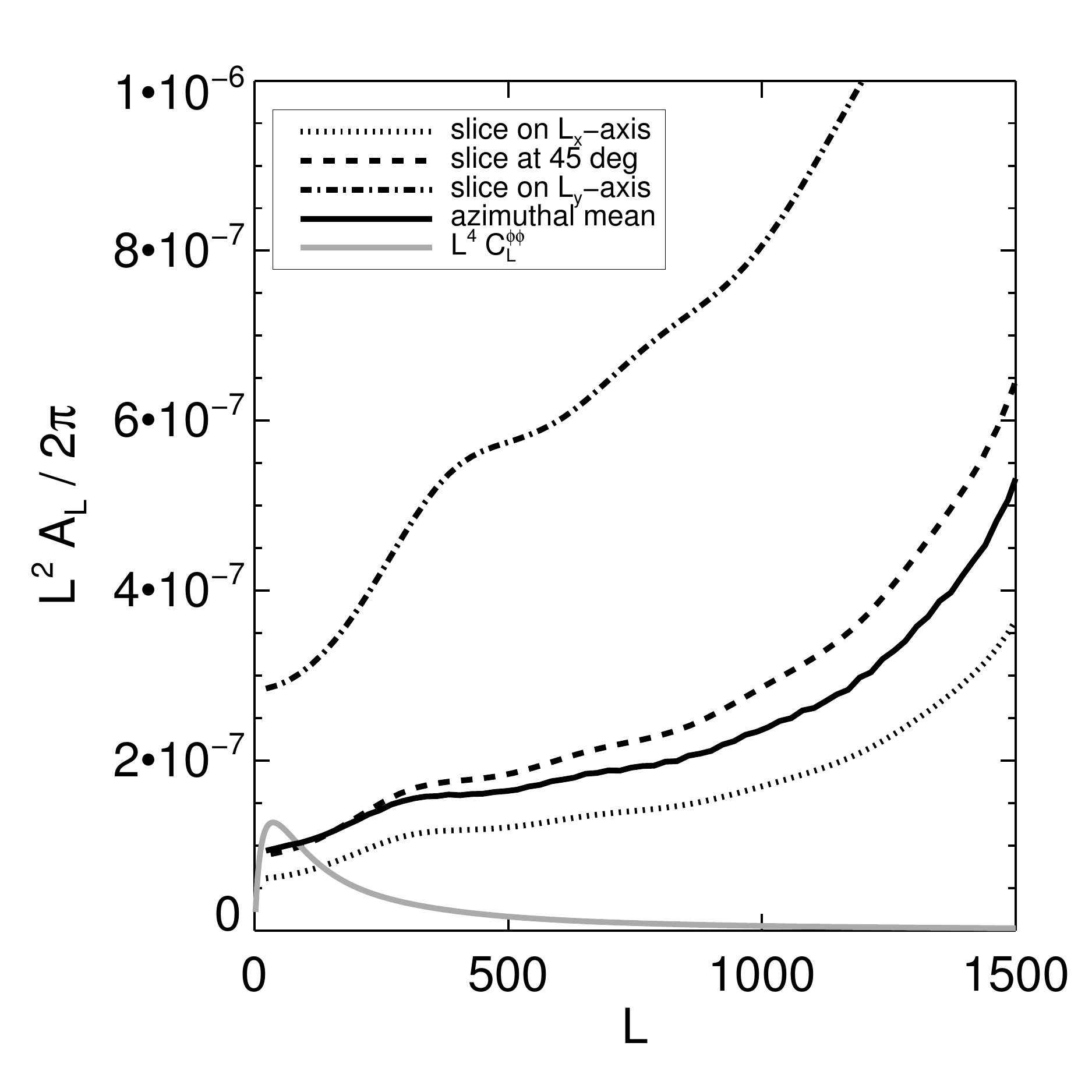}
\caption{Theoretical lensing reconstruction noise curves $A_{\Lvec}$
  (Eq.~\ref{eq:weighting}) for SPT map filtering and noise levels,
  showing slices along the $L_x$-direction (dotted), along the
  45-degree line (dashed) and in the $L_y$-direction (dot-dashed).
  Due to the anisotropic noise statistics and filtering, the lensing
   data are $\sim 4$ times noisier in the $L_y$-direction
  than in the $L_x$-direction.  The azimuthally-averaged mean
  theoretical noise curve, given by the thick solid line, is the noise
  bias which must be subtracted from the lensing power spectrum
  estimate.  Note that the variance of the bandpowers in the lensing power
  spectrum estimate will not rise as quickly with $L$, as the number
  of lensing modes to average over is $\propto L$.  The fiducial
  lensing power spectrum $L^4 C_L^{\phi\phi}$ is given by the grey
  line.}
\label{fig:ennellslices}
\end{figure}

Given a reconstructed map of the deflection field, $\dhat(\nhat)$, and associated analytic
function, $\al$, we construct a straightforward lensing power spectrum estimate
by averaging the map's Fourier amplitudes in azimuthal annuli.  We
down-weight the noisier lensing modes by applying an inverse-variance
weight $\propto (L^2 \clphi + \al)^{-2}$ in the azimuthal average. 
For $\clphi$, we use the lensing power spectrum predicted for the fiducial cosmology.  The anisotropic
theoretical noise level, $\al$, is shown as slices through the Fourier
domain in Figure~\ref{fig:ennellslices}.  Because of the filtering of 
$\ell_x < 1000$, as well as the anisotropic SPT noise power, the
lensing noise is highly anisotropic in the Fourier domain.

Neglecting covariance between bandpower estimates and assuming that
the filters are properly set, the reconstruction is expected to follow
\citep{hu01b, kesden03}
\begin{align} \label{eq:lensingvariance}
  \langle \dhat(\Lvec_1) \dhat(\Lvec_2 ) \rangle = & (2\pi)^2
\delta(\Lvec_1 + \Lvec_2) (L_1^2C_{L_1}^{\phi\phi} + N_{L}^{(0)}) \nonumber\\ & +
\mbox{(higher-order terms)}.
\end{align}

The sensitivity to the lensing power spectrum $\clphi$ originates from the connected part of the CMB four-point function, or trispectrum.  
The leading noise bias in the reconstruction, $\nlzero$, originates
from the unconnected, or purely Gaussian, part of the CMB four-point
function.  It is present even if lensing reconstruction is
performed on an unlensed CMB field \citep{hu01b, amblard04}.  Its theoretical expectation for a field without windowing is shown as the solid black line in Figure~\ref{fig:ennellslices}.  

For our analysis, we take two distinct  approaches in dealing with this bias:

\begin{itemize}
\item We use full SPT maps of each field to construct lensing maps,  and then estimate the lensing power spectrum of each map. This method has the most  statistical power, but has a noise bias that must be subtracted.
\item We split the SPT data into ``low-$\ell$'' and ``high-$\ell$''
  bands by applying low-pass and high-pass spatial filters.  We estimate two lensing deflection fields, $\dhat_{\rm low}(\Lvec)$ and $\dhat_{\rm high}(\Lvec)$, from the two bands.  We then compute the cross-correlation between these two lensing maps.  A lensing detection obtained with this approach will have lower signal-to-noise, but no  Gaussian noise bias.
\end{itemize}

\subsubsection{All-$\ell$ technique}
\label{sec:removen0}

In this approach, we calculate the expected noise bias for each field using the unlensed
end-to-end simulations of Section~\ref{sec:lenssims}. 
 The SPT temperature calibration could potentially vary between the four fields used in this analysis.  
 We calibrate each field by comparing the average temperature power spectrum in the $1200 \le \ell \le 3000$ range to the temperature power spectrum used in our simulations, which allows us to remove the Gaussian bias with high accuracy.  
 Each field is rescaled at the 1\% level in temperature, and we propagate the residual uncertainty in each field's calibration to the lensing power covariance matrix.

\subsubsection{\lsplittext\ technique}

An incorrectly-calculated noise bias will lead to an anomalous signal
in the measured lensing power spectrum.  An alternative to directly
characterizing and removing this bias is to construct two maps of
the same lensing field, using CMB maps with no modes in common
\citep{hu01b, sherwin10}.  This can be achieved, for example, by
filtering to isolate CMB multipole ranges in two disjoint annular
regions.  The estimated cross spectrum between these two reconstructed
lensing maps will then contain no Gaussian bias, since the maps have
no modes of Gaussian CMB,  instrumental noise, or foregrounds in common. 

However, due to the smaller number of mode pairs used to construct
each of the lensing maps, the resulting lensing power spectrum
estimate will have a significantly lower signal-to-noise ratio.  We
use a Fisher matrix approach to forecast the signal-to-noise as a
function of the split multipole $\lsplit$.  For the SPT noise and
filtering, we find that the highest possible detection significance
with this split is smaller than that in the all-$\ell$ analysis by a
factor of $\sim 0.38$, for the SPT noise and filtering.  We find that
this quantity is maximized when the split multipole is set to $\lsplit
\simeq 2300$.  This corresponds to a cut slightly higher than the
center of the main signal band, $1200 < \ell < 3000$.

For each field, we compute lensing maps using only CMB modes with
spatial frequencies either of $1200 \leq \ell \leq 2200$ or $2300 \leq
\ell \leq 4000$, with the gap of width $\delta \ell = 100 $ between
the two annuli being necessary due to the convolution by the finite
apodization window.  We estimate the lensing signal for each of these
maps, and construct a cross power spectrum.  We refer to
this as the ``\lsplittext'' technique for the remainder of the
paper.

\subsubsection{Higher-order biases}
\label{sec:biases}
There are known additional terms in Eq.~\ref{eq:lensingvariance} that
affect the reconstructed lensing power spectrum.  At high
$L$, there is a positive bias that arises from 
correlations in the CMB trispectrum generated by lensing
\citep{kesden03}.  This bias is proportional to the lensing signal,
$\clphi$, though evaluated at a different set of multipoles.  It is
denoted \nlone due to its linear dependence on the lensing power
spectrum.  This effect also leads to an excess in the power spectrum of
the estimated curl field as we show in Appendix~\ref{sec:curlappendix}.

There is an additional negative bias which arises at low $L$, due to
effects in the reconstruction of order $(\clphi)^2$ \citep{hu07, hanson11}.
This bias is denoted \nltwo due to its second-order dependence on the
lensing power.  This effect is neglected in the formulation of the
quadratic estimator, which only considers the lensing operation in the
map to linear order in $\phi$ (Eq.~\ref{eq:lensmodecoupling}).

In Section~\ref{sec:lensamp}, we present our detection of lensing in
terms of the excess signal compared to \nlzero, which is determined
from the unlensed simulations.
To calibrate the level of lensing power detected, we compare                                        
the excess power in the SPT measurements to the excess found in the                                  
same analysis of the lensed simulations.
Since the lensed and unlensed simulations contain equivalent amounts
of Gaussian, on-diagonal CMB power (with $\lvec_1 + \lvec_2 = 0$), we
label the excess signal seen in the lensed case as the signature of
lensing.  This simulation-based approach leads to a detection of the
effects of lensing which naturally takes the higher-order biases into
account, obviating the need the need to model these biases precisely.

\section{Systematic Uncertainties}

Uncertainties in the underlying cosmology and contributions from
foregrounds have been shown to potentially bias the reconstructed
lensing power spectrum \citep{cooray03,amblard04,perotto10}.  
In particular, \citet{amblard04} have shown that the thermal and kinetic
SZ effects can significantly bias the estimates.  
Here we re-evaluate
these biases,  particularly
given more recent measurements of the amplitude of the thermal SZ
power spectrum, Poisson point source power, and clustered point source
power at 150$\,$ GHz \citep{hall10, dunkley11,shirokoff11,reichardt11}.  We also study  the impact of non-linear structures in the Universe and beam uncertainties 
in the context of our analysis approach. 
  
We will show in the results section
(Section~\ref{sec:results}) that the lensing estimator with the
highest signal-to-noise ratio, namely the all-$\ell$ method, yields a
statistical error on the total lensing amplitude of 15\% when applied
to our data.  Biases on the reconstructed lensing power spectrum which
are substantially smaller than this quantity can be safely neglected
in our analysis. We show below that none of the possible sources of bias
that we consider in this section show evidence for being significant.

\subsection{Foregrounds}
\label{sec:foregrounds}

Emission from galaxies and Galactic dust are a possible source of
non-Gaussianity, and could in principle be a problem for CMB lensing
reconstructions. In particular, we investigate the impact of
infrared and radio galaxies, SZ effects, and Galactic cirrus. We will
show that these foregrounds are unlikely to be a substantial source of bias.

\subsubsection{Infrared and Radio Galaxies}

Point sources will affect the estimator in two ways: they
will add Gaussian power to the CMB map, and the brightest sources will
generate a trispectrum which will lead to an apparent lensing signal.
The purely Gaussian component is similar to the experimental noise, and
its presence will slightly raise the effective Gaussian noise bias. 
This is naturally taken into account by our handling of the noise bias
described in Section~\ref{sec:removen0} in the case of the all-$\ell$ analysis, and is not present as a bias in the \lsplit\ analysis.

The point source trispectrum is a potential contaminant.  Due to the
non-locality of the lensing estimator, a bright point source  will lead to spurious signal on
all scales.  Here, we determine at what flux level the map must be
cleaned of bright sources so the residucal Poisson point source background will be in the Gaussian limit.to put the Poisson point source background
into the Gaussian limit for the lensing estimation.

We simulate fields of Poisson-distributed point sources using number
count models for dusty, star-forming galaxies given by
\citet{negrello07}, and for radio sources from \citet{dezotti05}.  In the case of the dusty star-forming galaxies we
scale the counts to 150$\,$GHz using the same assumptions on the spectral indices
as those described in \citet{hall10}.  These counts agree, up to 100\,mJy, with the recent measurements of these populations at
150\,GHz \citep{vieira10,marriage11a,planck11-13}.  We generate randomly-placed
sources with flux values between 0.01 and 10$^3\,$mJy.

We pass these point source simulations, together with simulated lensed
CMB fields and noise realizations, through the lensing estimator.
Since we are specifically seeking to isolate the non-Gaussian
contribution of the Poisson point sources, for comparison we also pass purely
Gaussian fields  through the
estimator.  The Gaussian fields are constructed using the same power spectrum as the non-Gaussian foreground fields, including the effects of source masking.  For simplicity, in this section we use a version of the
estimator which is formulated for maps with periodic boundary
conditions, bypassing the extra apodization step.

We find that with a flux cut of $10\,$mJy, the uncorrelated trispectrum contribution from
Poisson sources is equal to that from equivalent Gaussian
power, for both the infrared and radio sources, to within 1\% in the reconstructed lensing power spectrum.  Turning the flux cut
up to $20\,$mJy, a $\sim 5$\% bias on the reconstructed lensing power
becomes apparent in both the all-$\ell$ and $\ell$-split
reconstructions.  Given that we remove sources in the 150$\,$ GHz SPT
maps at thresholds of $\sim 10\,$mJy, we conclude that
the non-Gaussian contribution of Poisson point sources are an
insignificant source of bias on the lensing reconstruction.

We also perform equivalent estimates for the curl signal in these
fields. These estimates show negligible signal, indicating that the curl estimate is not a useful check for Poisson foregrounds.

The angular fluctuations in the cosmic infrared background (CIB), as well as
the Sunyaev-Zel'dovich effects, are expected to be correlated with the
mass fluctuations responsible for CMB lensing \citep{song03,
  cooray00b}.  This is because these sources are tracers of the same
underlying three-dimensional matter field, and are thought to have a
similar distribution in redshift as the CMB lensing redshift kernel
(Eq.~\ref{eq:redshiftintegral}).  Clustering of the CIB sources has been detected at SPT wavelengths at high
significance \citep{hall10, hajian11, planck11-6.6_arxiv, shirokoff11, reichardt11}.  The CIB-$\phi$ correlation can potentially bias the
lensing estimate \citep{cooray03}. To test this possibility, we
use two separate lines of investigation: Gaussian random fields
that have a CIB field completely correlated with the lensing
convergence map, and the simulations of \citet{sehgal10}.

As a first test for a correlated signal, we assume that both the lensing convergence
and the CIB trace the linear density
fluctuations in the Universe and that both are Gaussian random fields.  
We assume that the CIB field is completely correlated with the convergence
field and normalize the amplitude to match the observations of 
\citet{reichardt11}.  In this case, we find no measurable bias on the lensing reconstruction.

We also perform an analysis on maps from the IR simulations by
\citet{sehgal10}.  Since the simulations were performed, much has been
learned about the mm-wave properties of the CIB; for example, these
simulations assumed a frequency scaling from 353 GHz down to 150 GHz
that was more shallow than has been observed \citep{reichardt11,
  addison11}. Using a more appropriate frequency scaling of the dust
emissivity than that assumed for the \citet{sehgal10} simulations
leads to the CIB maps being reduced in amplitude by a factor of
1.7. Scaling the maps by this factor leads to a power spectrum from
Poisson-distributed dusty sources that is in excellent agreement with
\citet{reichardt11}.  

To study the nature of biases from the Sehgal et al.\ CIB simulations,
we rotate the CIB fields by $90^\circ$ to break the correlations
between the CIB and lensing fields.  We find a small bias ($<1\%$ at
$L=500$) from the CIB sources in the absence of these correlations.
Restoring these correlations, the bias in the lensing power spectrum
at $L<500$ is found to be $\sim -$3--4\%, which is smaller than the
statistical uncertainty in our analysis (detailed in
Section~\ref{sec:results}).

A complete understanding of the impact of correlations between CIB
fluctuations and lensing convergence remains to be determined. The
contrast between the results from purely Gaussian simulations (showing
no contamination) and the Sehgal simulations ($\sim -3-4\%$ at low L)
demonstrate that careful CIB modeling will be required for future
analyses.  The Sehgal et al.\ simulations, while useful for these
purposes, have features which make them difficult to interpret. For
example, the source counts are lower than observations at 150\,GHz
\citep{vieira10} between 5 and 10\,mJy, which is close to the flux cut
that we employ; the amplitude of the CIB power spectrum from clustered
sources is also lower than that seen in recent measurements; and the
finite simulation volume (1 Gpc/$h$) subtends only 25 degrees at z=1.
Larger simulations, created with input from recent observations at mm
wavelengths, should help gain a better understanding of this systematic effect.

\subsubsection{Thermal and Kinetic Sunyaev-Zel'dovich Effects}

As with the radio and infrared sources, the temperature
decrement associated with an unmasked massive SZ cluster leads to a
large feature in the reconstructed deflection map.  This feature can
potentially generate a bias on the reconstructed lensing power
spectrum, and correlations of the SZ field with the lensing field could
lead to a bias in the observed lensing power 
\citep{cooray03}.

Our simulated SPT observations, described in
Section~\ref{sec:lenssims}, contain Gaussian fluctuations with an SZ
power spectrum template.  However, they do not contain discrete SZ
clusters.  Unlike the case for the radio and infrared galaxies, the
masking of SZ clusters in the SPT maps does not have an equivalent
procedure in our simulated observations.  Masking of objects in the
data but not in the simulations leads to a small difference 
in the amount of temperature power in the maps.

Given the possible bias from the thermal SZ signal,
we conduct three analyses to assess its importance using two independent
thermal SZ simulations, along with an empirical measurement of the importance
of SZ masking in the data analysis.

We use the maps of \citet{sehgal10}, rescaling the amplitude of
these maps to match the lower SZ power spectrum seen in
measurements \citep{lueker10, fowler10,reichardt11} which were made after these
simulations were created.  As with the radio and infrared sources, we
pass these simulations through the lensing estimator, together with
simulated CMB and noise fluctuations.

In each case, we compare with a Gaussian field with equivalent power
spectrum.  We identify two sources of bias from the thermal SZ effect
in the Sehgal et al.~simulations. The very bright objects contribute a
large enough signal in the lensing reconstructions to add a positive
bias of $\sim 20\%$ to the lensing power spectrum without any masking,
while correlations of massive galaxy clusters with the large scale
structure responsible for lensing lead to a negative bias of $\sim 10\%$.  Masking of
the extremely bright SZ clusters, as is done in the data, reduces the
positive bias from the most massive clusters to be less than 10\% of
the lensing power spectrum, and the negative bias from correlations
with large scale structure reduced the total bias from the thermal SZ
effect to be negligible for this analysis ($<3\%$).

The second simulation that we use as an independent test is a 500 Mpc/h
N-body+SPH simulation with $1024^3$ dark matter as well as gas particles, that is performed with a different assumed
cosmology \citep{gottloeber07}.
Snapshots are approximately sampled at each light-crossing time.  We then produce maps by randomly rotating and translating each simulation volume and creating an SZ map, then ray tracing over all data cubes adding in each case the SZ effects at the deflected position. 
After rescaling the maps to agree with the
observed SZ power spectrum, we pass these maps through the lensing
estimator.  In this case, a large bias (50\% of the lensing power
spectrum at $L=700$) is generated by the most massive clusters in the
map, but masking of the very brightest SZ sources again reduces the bias in the
lensing power spectrum to negligible ($\sim 5\%$) levels.

Given theoretical uncertainties associated with simulations of the thermal SZ \citep{dunkley11, reichardt11} we also take an empirical approach
by running the full lensing reconstruction pipeline on SPT maps with
 differing masking levels.  We find that without any masking of
clusters, the best-fit lensing amplitude increases by $0.12\,\sigma$ (2\% of the lensing power spectrum amplitude)
compared to the result, presented below, which contains $0.01$
clusters masked per square degree.  At the more aggressive masking
level of $0.07$ clusters masked per square degree, the best-fit
lensing amplitude decreases by an equivalent amount, $0.12\,\sigma$.
We therefore conclude that thermal SZ is not a substantial source of
bias in this analysis.

We also run simulated kinetic Sunyaev-Zel'dovich fields through
the lensing estimator. Again we repeat this analysis for the Sehgal. et al. as well as SPH simulations.  The power spectrum of these fields is
consistent with current upper limits \citep{reichardt11}.  This leads
to a bias that is equivalent to a fully Gaussian field with the same
power spectrum, to within 1\%.

\subsubsection{IR cirrus}

Diffuse Galactic dust emission is known to be an important foreground
for CMB studies. The SPT fields are  chosen to minimize Galactic emission, but cirrus emission is detected at $\sim$\,3\,$\sigma$ through cross-correlation with  maps from the Infrared Astronomical Satellite \citep{finkbeiner99}, as described in K11.
To test the importance of this
cirrus contamination, we subtract a template based on
\citet{finkbeiner99} from the field that shows the strongest cirrus
detection (the \five\ field), and re-calculate the lensing power
spectrum.  In no $\ell$-bin is the result changed by more than 2\%,
and there is no evidence for a systematic bias. The non-Gaussianity of
the Galactic cirrus in our fields is not a serious contaminant for CMB lensing
studies at 150\,GHz.

\subsection{Uncertainty in the unlensed CMB temperature field}
\label{sec:unlensedtemp}

Even if the Gaussian noise bias is perfectly removed, the lensing map
is subject to a calibration uncertainty which arises from uncertainty in the 
CMB power spectrum, as mentioned in \citet{hu01b}.  This is because the
lensing estimate is based on the mode-coupling of
Eq.~\ref{eq:lensmodecoupling}; an uncertainty in the unlensed CMB
power spectrum will lead to a multiplicative offset on the
reconstructed lensing map.  

In this analysis, we enforce a constraint that the power spectrum of
the data match that of the simulations.  For a fixed theoretical 
lensed CMB power spectrum, the uncertainty in the theoretical unlensed CMB
power spectrum is small. Sample variance between lensed and unlensed
power spectra for a given realization are strongly correlated, but in
any case would be less than 1\% in amplitude for the sky coverage and
$\ell$ range considered.  Uncertainty in the CMB power spectrum is therefore not a 
limitation of this analysis.

The effects of primordial non-Gaussianities upon the estimator have
been shown to be negligible, being two orders of magnitude smaller
than the first-order bias of \citet{kesden03} for non-Gaussianity
parameters $f_{NL}$ and $g_{NL}$ consistent with current upper limits
\citep{lesgourgues05}.

\subsection{Effects from nonlinear growth of structure}
\label{sec:nonlinear}
Interaction between the angular scales of the gradient and lens 
(Eq.~\ref{eq:filter}) can lead to a negative bias of the lensing reconstruction
\citep{hu07, hanson11}. This effect is expected to increase in the
presence of nonlinear structures like clusters and filaments. 
To determine whether our simulation pipeline using Gaussian random fields 
of matter fluctuations (described in Section~\ref{sec:lenssims}) leads to an unbiased reconstruction, 
we run a cosmological N-body simulation of gravitationally interacting 
dark matter particles, using the Gadget-2 code.\footnote{http://www.mpa-garching.mpg.de/gadget/right.html}
The simulated cube is 1000 Mpc/h on each side
with $1024^3$ particles. We again produce ray-tracing simulations of CMB lensing in maps $15 \times 15$ degrees on a side that start from a regular grid near the observer and deflect each ray at the interpolated positions at each projected plane. The resulting lensing potential has excess small scale power as expected in nonlinear structure formation, and the lensed temperature power spectrum is found to be consistent with the nonlinear lensing option of CAMB \citep{lewis99} out to $\ell \simeq 5000$. This confirms convergence of the resolution of our dark matter simulation on all scales relevant for the SPT lensing reconstruction.
The lensing estimator (Section~\ref{sec:theorybackground}) is then
applied to the lensed maps and the reconstructed lensing power
spectrum compared to that of the input. The procedure is repeated for
100 maps produced from different randomly chosen translations and
rotations of the simulation volumes along the line of sight. 
We find a deviation of the second-order lensing bias $\nltwo$ from the
Gaussian case with magnitude  equal to a few percent of the total
reconstructed power.  This bias is insignificant compared to sample
variance for the sizes of the SPT fields.

\begin{figure*}[t!]
\epsscale{1.2}
\vspace{-2cm}
\plotone{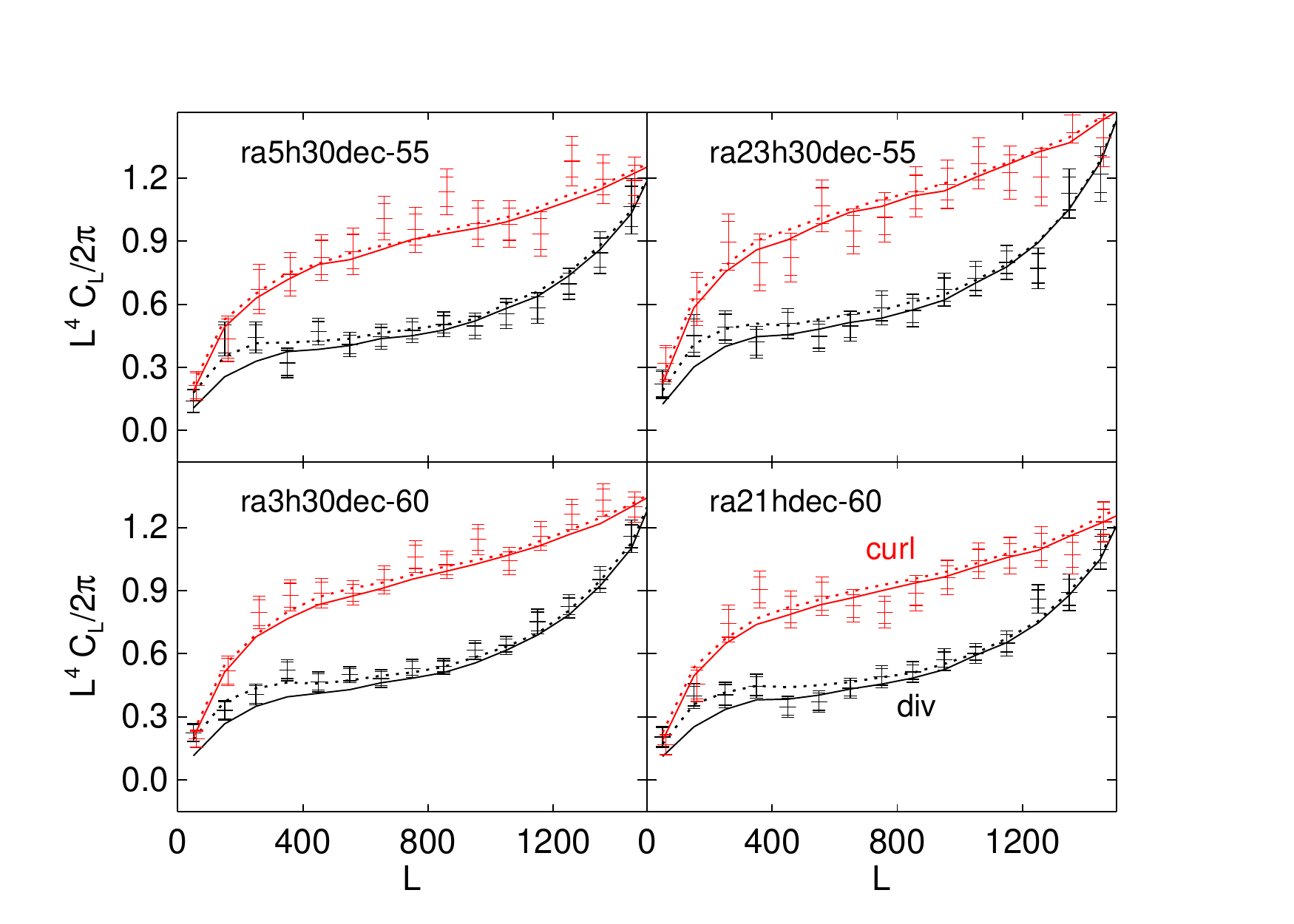}
\caption{Individual all-$\ell$ raw power spectra for each field for the main lensing
signal (the divergence; bottom, black, points in each panel) and the curl component
(top, red, points in each panel). Curves show the results of the lensed and unlensed simulations; i.e.,~the lower curves show the Gaussian noise biases estimated from
simulations and the upper curves show the sum of the noise bias and the expected lensing signal in
our fiducial cosmological model. The extra ticks on the error bars show the
impact of the correlated covariance arising from the uncertainty in the Gaussian noise bias subtraction.}
\label{fig:div_curl}
\end{figure*}

\subsection{Beam Uncertainties}
\label{sec:beamuncert}

Here we address the uncertainty in our analysis due to the uncertainty
in the SPT beam profiles.  The simulated observations convolve the sky
by the SPT beams.  If the beam used in this convolution differs from
the true SPT beam, the beam-convolved sky power, and thus $\nlzero$
bias, will differ between the data and the simulations.  This would
result in a bias in our $\nlzero$-subtracted $\clphi$.

To first order, this effect is removed when we recalibrate the data
maps such that their average beam-convolved temperature power spectra
are equal to the simulated beam-convolved power spectra.  However,
there is a residual uncertainty due to the tilt of the beam uncertainty
across the $1200 < \ell < 3000$ range.  We have checked that the
effect of this tilt is small.  If we repeat the analysis using
simulated beams that differ from the nominal beams by a 1\,$\sigma$ beam
uncertainty, we find that the best-fit lensing amplitude shifts only by -0.4\%, or
-0.03\,$\sigma$.  We conclude that the uncertainty in the beam has a
negligible effect on this analysis.

\section{Results}
\label{sec:results}

Two types of results are reported below.
First, the amplitude of the lensing signal is compared with expectations from
our simulations,  which are performed at a single point in
cosmological parameter space. 
We then explore the cosmological parameter space allowed by current
cosmological probes, using the lensing data to both better constrain
cosmological parameters and characterize the amplitude of gravitational
lensing compared to expectations from the ensemble of allowed cosmological
models.

\subsection{Measuring the lensing amplitude at a reference \LCDM cosmology}
\label{sec:lensamp}

The raw, unnormalized  power spectra of the estimated deflection maps are shown in Figure~\ref{fig:div_curl}.  The  spectra are  dominated by the lowest-order
noise bias, and, additionally have not been corrected for the effects of the
windowing or the higher-order biases. The noise bias is substantial, but it is also clear that the SPT data show an 
excess in all fields over the unlensed prediction. 
The curl estimator also shows a preference for
lensing, and demonstrates that the Gaussian noise in the SPT maps is
well-understood.

\begin{figure*}
\plottwo{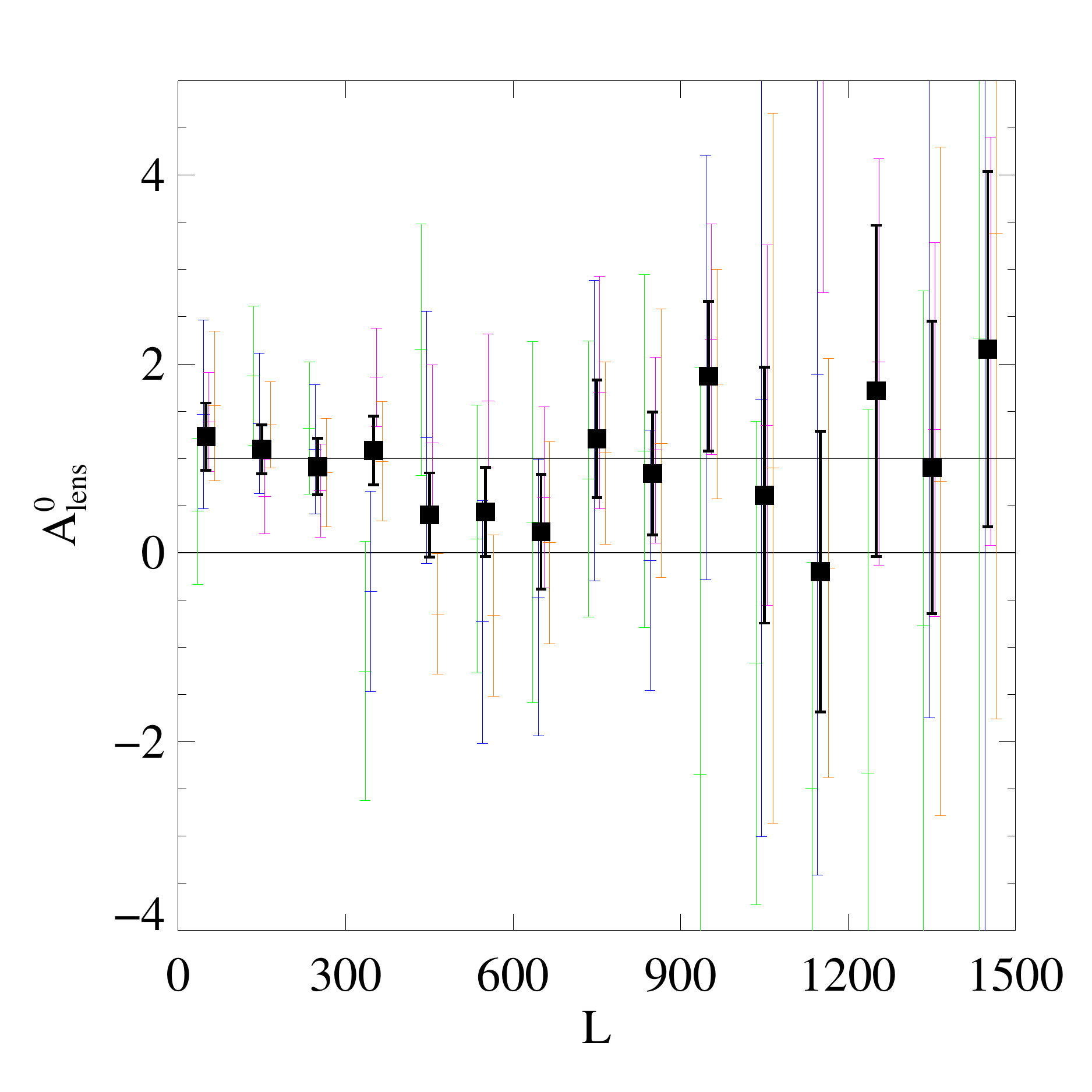}{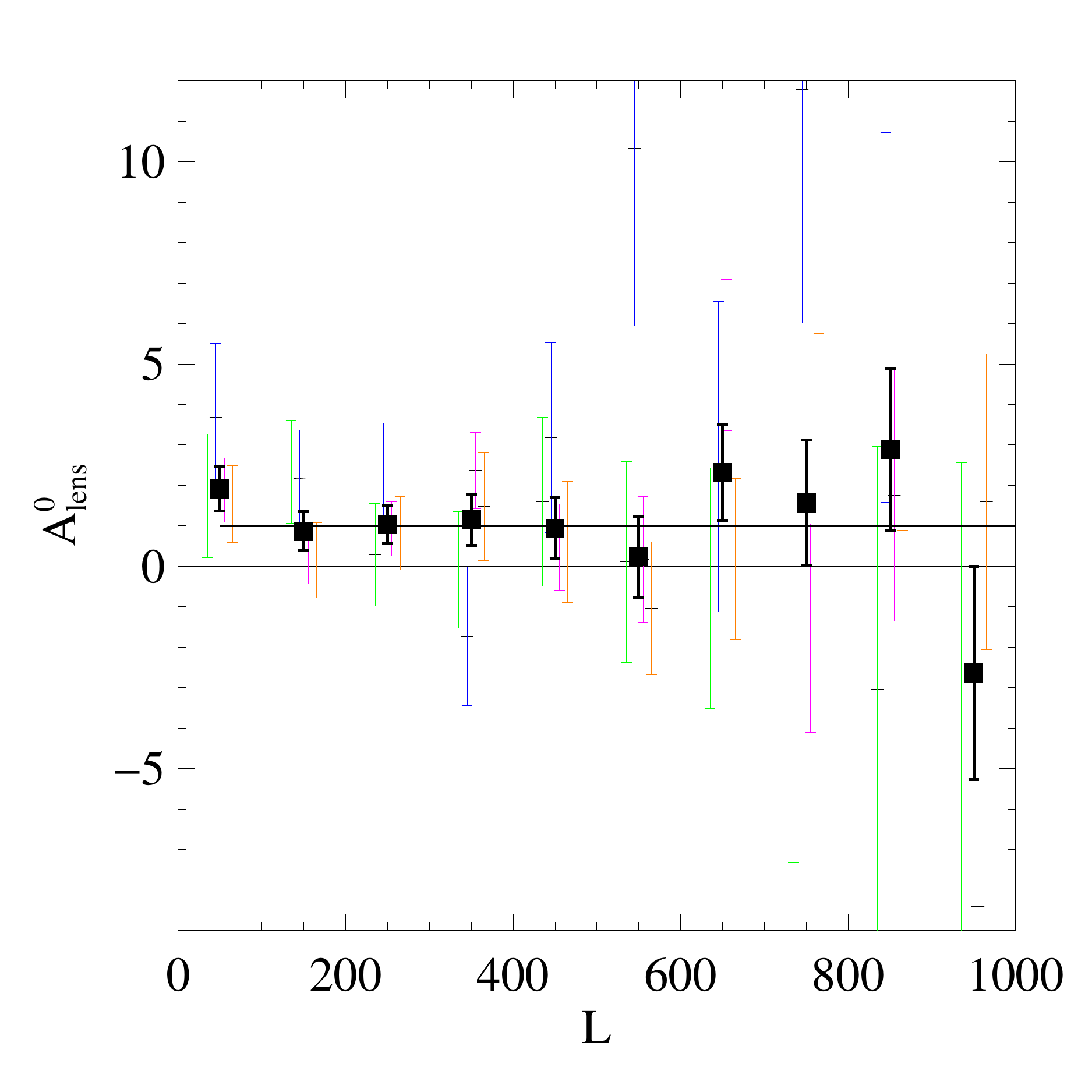}
\caption{Ratio of the power excess measured in the SPT data compared to the power excess from
lensed simulations. The left panel uses all-$\ell$ maps, while the right panel uses
the $\ell$-split method of using disjoint annuli in $\ell$ space to avoid a
noise bias. The horizontal lines indicate lensing amplitudes of 0 and 1.  Each field is shown as a different color, offset in $L$ for clarity: the \five\ field in green; the \twthree\ field in blue; the \three\ field in magenta; and the \twonesix\ field in orange.  The heavy black 
points show the combined best-fit estimate of the lensing amplitude.
Note the expanded scale in the right panel; the \lsplittext\ method has less statistical power.  No lensing is excluded at 6.3$\,\sigma$ (left) and $3.9\,\sigma$ (right)}
\label{fig:alens_L}
\end{figure*}

After subtracting the expected noise bias, we compare the measured excess in the divergence estimator to that seen in the lensed simulations in the left panel of Figure~\ref{fig:alens_L}.   The relative bandpowers are shown as an $L$-dependent scale factor
\begin{equation}
\alenszero(L) = \frac{\hat{C}_L^{\rm data} - \hat{N}_L^{(0)}}{\hat{C}_L^{\rm sim} - \hat{N}_L^{(0)}}. 
\end{equation}
Here, $\hat{C}_L^{\rm data}$ is the raw power spectrum of the reconstructed lensing deflection field; i$\hat{C}_L^{\rm sim}$ is the field-dependent raw power spectrum of the lensed simulations; and  $\hat{N}_L^{(0)}$ is the field-dependent noise bias, which is obtained by performing equivalent reconstructions on the unlensed simulations.  The superscript (0) refers to this being the amplitude of the lensing signal relative to the template provided by our simulations.
With this definition, $\alenszero(L) =1$  corresponds to the amplitude of the lensing signal in the simulations; $\alenszero(L)=0$ corresponds to no lensing signal.  We show this quantity for the all-$\ell$ analysis in the left panel of Figure~\ref{fig:alens_L}; evidence for lensing can clearly be seen.   The right panel of Figure~\ref{fig:alens_L} shows the same quantity for the $\ell$-split technique, which has no Gaussian noise bias needing removal.

We then fit the measured $\alenszero(L)$ to the model of an $L$-independent lensing amplitude $\alenszero$ which scales the amplitude of the lensing power spectrum in our fiducial cosmology.  
We assume a Gaussian likelihood function of the form 
\begin{align}
\label{eq:alenslike}
-2 \ln\mathcal{L}(\alenszero)= & \ln \det(\mathbf{C})   + \\ \nonumber & \sum_{LL^\prime} (\alenszero(L) - \alenszero) \\ \nonumber
&\times \mathbf{C}_{LL^\prime}^{-1} (\alenszero(L^\prime) - \alenszero)
\end{align}
We obtain an approximation to the bandpower covariance matrix $\mathbf{C}$ using 2000 lensed flat-sky simulations which include apodization  for the \five field.  This large number of simulations is necessary due to the large scatter in the off-diagonal terms.  The bands are correlated at the 15--20\% level;  the shape  of the off-diagonal elements in the covariance matrix is found to be similar to that obtained by \citet{kesden03} and \citet{hanson11},   using mode-counting arguments.  We additionally account for sample variance in the lensing amplitude  by scaling the diagonals of the covariance matrix by a factor $\propto [(\alenszero\clphi + \nlzero)/(\clphi + \nlzero)]^2.$   

To test the assumption of Gaussianity in the likelihood function, we also compare with  an offset-lognormal likelihood function \citep{bond00}.  We find equivalence with the two approaches in both the best-fit point and the width of the likelihood curves.

\begin{figure}
\plotone{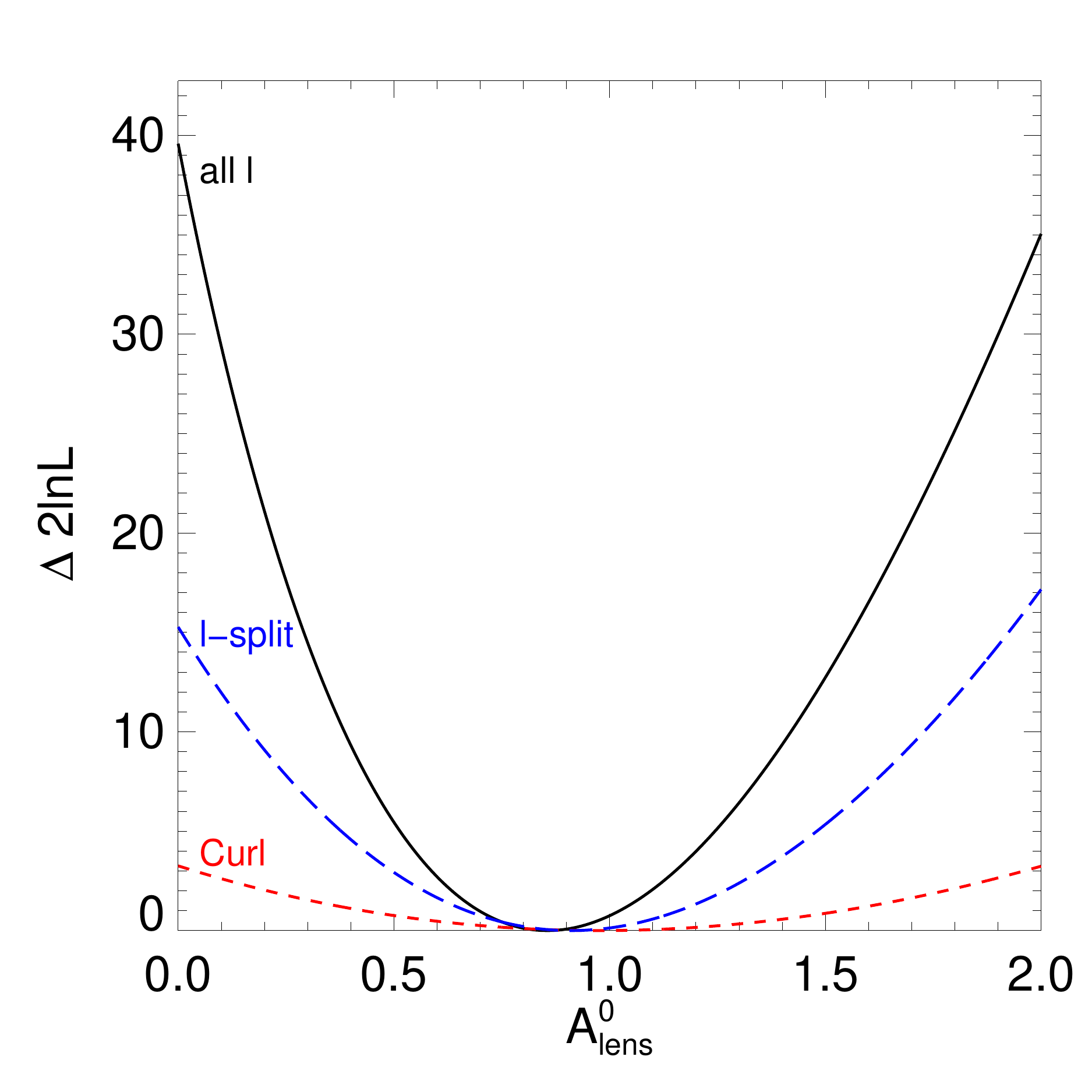}
\caption{Change in $2 \ln L$  compared to best fit for the SPT
  lensing power spectrum,  when the fiducial lensing power spectrum is multiplied by a lensing scale factor $\alenszero$. A
  strong detection is evidenced for both the less-sensitive \lsplittext method
 (blue, long dashed line) and the
  more-sensitive all-$\ell$ technique (black, solid line). 
  Using the curl signal in the data (red, short dashed), 
  lensing is also tentatively detected.}
\label{fig:chisq}
\end{figure}

Figure~\ref{fig:chisq}  shows the total likelihood for the fields as a function of $\alenszero$ and indicates a robust detection of
lensing power. No lensing is excluded at $3.9\sigma$ using the $\ell$-split approach, and at
$6.3 \sigma$ using the all-$\ell$ approach.  These quantities are quoted in terms of the difference in the likelihood function between 0 and the best-fit \alenszero, taking the total likelihood as the sum of likelihoods for the individual fields.
Using
the divergence signal in the all-$\ell$ maps, the best-fit lensing
amplitude is found to be $0.86 \pm 0.16$. A substantial component of this uncertainty
comes from the uncertainty associated with the \nlzero removal; in
the absence of this uncertainty, the error bar would be $\pm 0.11$.

\begin{table*}
\begin{center}
\begin{threeparttable}
\caption{$\chi^2$ values  and maximum likelihood fits for each field }
\begin{tabular}{l|ccc|ccc|ccc}
\multicolumn{1}{c}{} 
& \multicolumn{3}{c}{$\chi^2 (\alenszero=1)$} & \multicolumn{3}{c}{$\chi^2(\alenszero=0)$} & \multicolumn{3}{c}{Best fit \alenszero} \\
Field Name   & Div (all-$\ell$) & Div (\lsplittext) & Curl & Div (all-$\ell$) & Div (\lsplittext) & Curl & Div (all-$\ell$) & Div (\lsplittext) & Curl \\
\hline
\five & 12.0 & 4.0& 18.7 & 31.0 & 5.1 & 20.0 & $1.40 \pm 0.45 $ & $0.58\pm 0.60$& $1.1^{+1.3}_{-1.1}$\\
\twthree & 14.3 & 14.3 & 12.0 & 21.7 & 22.4 & 12.3 &$0.77 \pm 0.39$& $1.92\pm 0.69$&$0.2^{+1.3}_{-0.2}$ \\
\three & 10.6 & 13.9 & 13.4 & 25.2 & 20.6 & 17.9 &$0.84 \pm 0.28$&$0.90\pm 0.38$& $2.0^{+0.9}_{-0.9}$\\
\twonesix & 22.5 & 5.1 & 16.3 & 31.3 & 6.7 & 16.3 &$0.63 \pm 0.26$&$0.60\pm0.47$& $0.1^{+1.0}_{-0.1}$ \\
\hline
Total  ( \# pts) & 59.4 (56) & 37.2 (36) & 60.4 (56) & 109.3 (56) & 54.8 (36) & 66.6 (56) & $0.86 \pm 0.16$ & $0.91\pm 0.25$& $0.98 \pm 0.55$\\
\end{tabular}
\label{tab:chi2} 
\begin{tablenotes} \item  The field-by-field $\chi^2$ values, with each
all-$\ell$ spectrum consisting of 14 points and each $\ell$-split spectrum
having 9 points, together with the field-by-field best-fit $\alenszero$ for
the different spectra. The curl uncertainties are asymmetric within each
field because we have assumed $\alenszero \ge 0$.
\end{tablenotes}
\end{threeparttable}
\end{center}
\end{table*}

The values of $\chi^2$, the second term in Eq.~\ref{eq:alenslike}, are
shown in Table~\ref{tab:chi2} for the individual fields.  For
$\alenszero=1$ (not the best fit) the highest $\chi^2$ value for any field
still has a 7\% probability of observing a higher value, and the total
$\chi^2$ for all 56 points has a probability of 35\% of observing a
higher value.  In contrast, all of the fields have higher $\chi^2$ for
a model with no gravitational lensing, and the sum of the fields has a
$\chi^2$ with a probability of observing a higher value of $8\times
10^{-5}$.

The $\ell$-split technique shows a clear lensing detection with
$\alenszero=0.91 \pm 0.25$. As expected, the signal to noise is 
substantially lower, but recall that this lensing power spectrum
does not suffer from the issues of noise bias that present
a challenge for the result that uses the all-$\ell$ maps.

Using only the curl signal in the all-$\ell$ maps, we find a tentative detection
of lensing: $\alenszero=0.98 \pm 0.55$. The lensing signal in this mode is
due to the equivalent of the \nlone bias mentioned above and discussed
in more detail in the Appendix.

\begin{figure*}
  \centering
\vspace{-1cm}
  \plotone{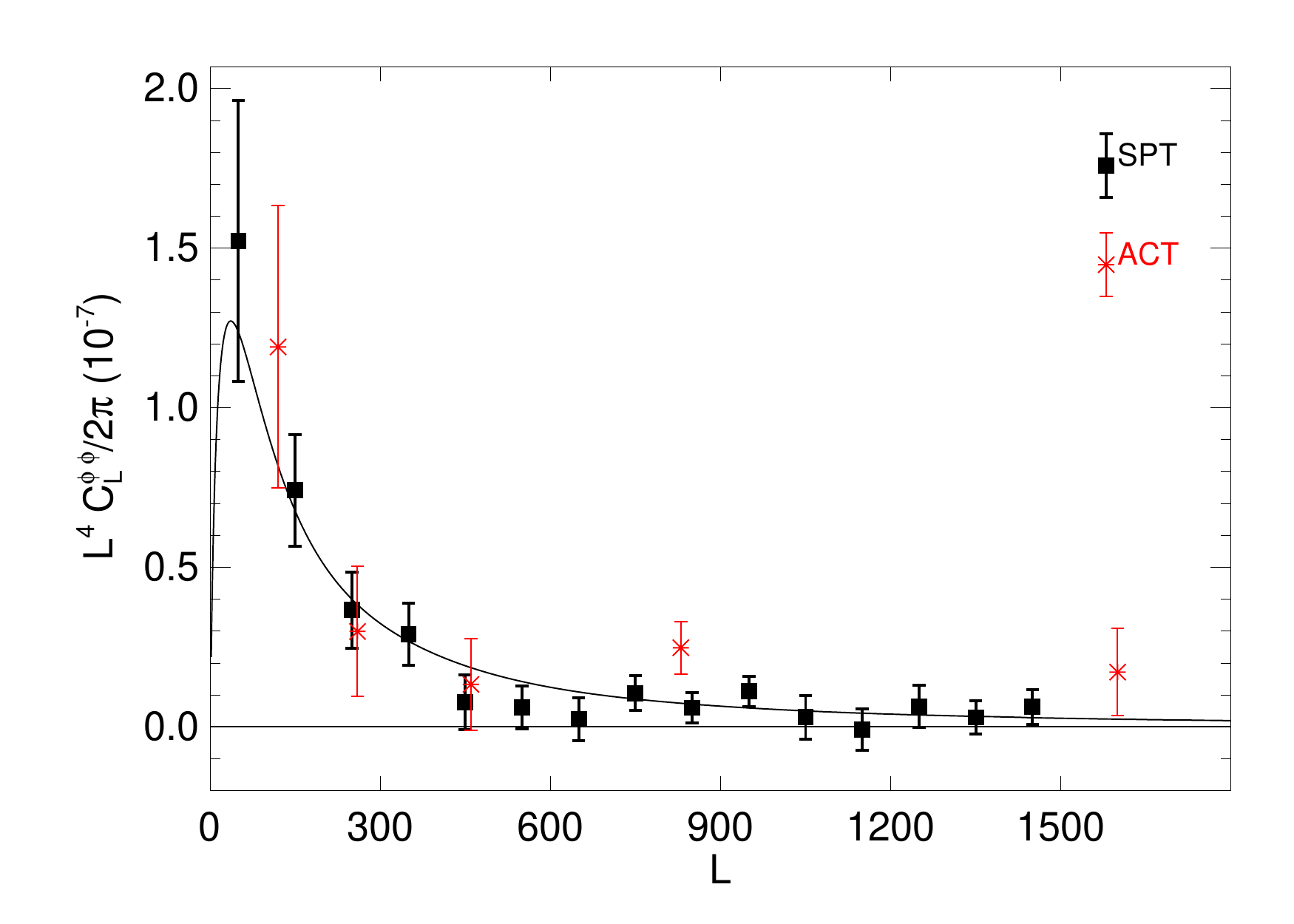}
  \begin{centering}
    \caption{A comparison of the derived lensing bandpowers from SPT and ACT \citep{das11}.  Although we show the lowest-$L$ datapoint, centered at $L=50$, we do not use this point in our fits due to the possible interaction with the subtraction of the apodization feature (Section~\ref{sec:apodization}) on this large scale.  The solid curve is not a fit to the data; rather, it is the lensing power spectrum in our fiducial \LCDM cosmology, corresponding to $\alenszero=1$.}
    \label{fig:cl_compilation}
  \end{centering}
\end{figure*}

Figure~\ref{fig:cl_compilation} shows the product of the derived lensing amplitudes as a function of $L$ and the  reference lensing power spectrum used in our simulations, $\alenszero(L) \clphi$.  This represents our best estimate of the lensing power spectrum.  These bandpowers are also shown in Table~\ref{tab:spt_bandpowers}.
 The \nlone and \nltwo biases lead to
non-local distortions of the lensing power spectrum.  However, if the
shape and amplitude of the true lensing power spectrum is similar to that in the 
assumed power spectrum, the true biases will not be significantly
different than what is assumed.

\begin{table}
\begin{center}
\caption[SPT lensing bandpowers]{SPT lensing bandpowers}
\begin{threeparttable}
\begin{tabular}{|c|c|c|}
\hline
$L$ & $L^4 \clphi / 2 \pi / 10^{-7} $&$ \sigma(L^4 \clphi / 2 \pi / 10^{-7})$\\
\hline
$  150 $&$  0.741 $&$  0.175$\\
$  250 $&$  0.366 $&$  0.119$\\
$  350 $&$  0.291 $&$  0.098$\\
$  450 $&$  0.077 $&$  0.085$\\
$  550 $&$  0.062 $&$  0.067$\\
$  650 $&$  0.025 $&$  0.067$\\
$  750 $&$  0.106 $&$  0.054$\\
$  850 $&$  0.060 $&$  0.047$\\
$  950 $&$  0.112 $&$  0.047$\\
$ 1050 $&$  0.031 $&$  0.068$\\
$ 1150 $&$ -0.009 $&$  0.064$\\
$ 1250 $&$  0.065 $&$  0.066$\\
$ 1350 $&$  0.030 $&$  0.052$\\
$ 1450 $&$  0.063 $&$  0.055$\\
\hline
\end{tabular}
\label{tab:spt_bandpowers}
\begin{tablenotes}
\item The lensing bandpowers, as shown in Figure \ref{fig:cl_compilation}.  Each value represents a band of width $\Delta L = 100$, centered at the given value of $L$. The final column shows the error within the given band, obtained from simulations.  The bandpowers of the lensing convergence $\kappa$ are related to those of the potential $\phi$ according to $C_L^{\kappa \kappa} = \frac{1}{4} L^4 C_L^{\phi\phi}.$  
\end{tablenotes}
\end{threeparttable}
\end{center}
\end{table}

\subsection{Cosmological Parameter Estimation with Extra Information from Lensing}

The power spectrum of the CMB lensing potential is sensitive to total matter fluctuations over a wide redshift range (peaking at  $z \sim 2$), and mainly on scales which are in the linear regime ($k \sim 0.05\, h/$Mpc).  A measurement of this power spectrum can therefore constrain physics which affects growth on these scales, as well as provide a distance measure to these redshifts \citep{kaplinghat03, smith06b, lesgourgues06, deputter09}.  The first cosmological constraints from the lensing power spectrum were produced with the ACT lensing reconstruction by \citet{sherwin11}.  When combined with data from WMAP, the lensing data showed a preference for spatial flatness,  and found a nonzero dark energy density at  3.2\,$\sigma$.
We first discuss the  constraining power from the SPT data in the \LCDM parameter space.  We then quantify the improvement in four additional parameters which we allow to vary: the amplitude of the lensing signal $\alens$;  the spatial curvature of the Universe $\Omega_k$; a nonzero sum of neutrino masses $\sumnu$; and the dark energy equation of state parameter $w$.

One complicating factor in using lensing measurements to obtain precision cosmological constraints is the nontrivial scaling of the higher-order biases with cosmological parameters \citep{kesden03, amblard04, hanson11}.  

In our approach, we obtain an effective $L$-dependent lensing amplitude $\alenszero(L)$ seen in the data relative to its expectation from simulations  (as shown in Figure~\ref{fig:alens_L}).   We then multiply by the fiducial lensing power spectrum $\clphi$ used in the simulations.  If the true cosmological parameters were exactly equal to those assumed in our simulations, then the higher-order biases would be  completely accounted for in this approach.  To consider different cosmological parameters, we must therefore estimate the different scalings with cosmological parameters for these higher-order effects.

The leading-order bias $\nlone$  at a given multipole $L$ is given as an integration over the lensing power spectrum at other multipoles.  
A parameter which scales the amplitude of the lensing signal, such as the scalar spectral amplitude $A_s$, will thus affect $\clphi$ and $\nlone$ in the same way.  
However, a parameter which affects the shape of the spectrum in a nontrivial way will affect $\clphi$ and $\nlone$ differently \citep{hanson11}.  
We note that the measurements at low $L$, which contain the highest signal-to-noise ratio, are dominated by  $\clphi$ (after the subtraction of $\nlzero$).  
The  size of  $\nlone$ becomes $\gtrsim 50\%$ of the signal  at $L = 1000$; however, the signal-to-noise ratio per band also decreases at high $L$.

To estimate the impact of the scaling of $\nlone$ with parameters, we analytically compute $\nlone$ at a grid of points in parameter space.  The calculation of the \nlone\ bias is a CPU-intensive four-dimensional integral in the Fourier domain for each value of $L$ considered.  For simplicity, we therefore compute \nlone assuming isotropic noise fluctuations.  This allows us to evaluate \nlone on a one-dimensional line $L$, rather than at a two-dimensional grid of points as would be necessary if considering the anisotropic SPT noise.  We then  numerically evaluate the derivatives
\begin{equation}
B_\alpha \equiv \left |{d \over d p_\alpha}  \ln\left({m(L)^2 L^4 \clphi +  L^2 \nlone} \over {m(L)^2 L^4 \clphi} \right )  \right | \sigma_{p_\alpha}. 
\label{eq:ennonescaling}
\end{equation}
The factor $m(L)^2$ encapsulates the  calibration offset in the lensing estimate due to uncertainty in the unlensed CMB power spectrum, discussed in Section~\ref{sec:unlensedtemp}; it is equal to unity if the unlensed CMB power spectrum is equal to its assumed value.  The set of parameters  $p_\alpha$  which we vary consists of $ (\Omega_bh^2, \Omega_ch^2, H_0,\tau,A_s, n_s, \Sigma m_\nu, \Omega_k, w)$.
As the final step, we multiply by the cosmologically-allowed $1\,\sigma$ range in the given parameter, $\sigma_{p_\alpha}$.  We find that the logarithmic derivative $B_\alpha$ is less than 0.02 for $L < 1300 $ for all parameters considered.  The two bins at higher $L$ constitute only 3.2\% of the total SPT lensing $\chi^2$, and contain effectively negligible weight in parameter fits.    $\nlone$ can thus be treated as a transfer-function effect on the lensing modes used in the current analysis.

The second-order, negative bias, $\nltwo$, appears on the largest
scales.  The SPT lensing bandpowers correspond to scales smaller than
a full-sky experiment, such as \emph{Planck}.  Using simulations, we
find that at $L=150$, the lowest $L$ at which we report our results,
its value is approximately $|N_{150}^{(2)}| = 0.60\,\sigma_{150}$.
Here, $\sigma_{150}$ denotes the uncertainty in the reported band at
$L=150$.  At $L=250$, its value is $|N_{250}^{(2)}| =
0.17\,\sigma_{250}$.  In our approach, the majority of the effect of
\nltwo is removed by scaling to its value in the reference cosmology.
Since the lensing amplitude is measured at the $\sim 20\%$ level from
the other bands, the uncertainty in this rescaling is small (on the
order of $\sim 0.2 \times 0.6\sigma_{150} = 0.12\,\sigma_{150}$ for
the band at $L = 150$, and smaller at higher multipoles).  We thus
neglect the effect of $\nltwo$ in the following analysis.

To explore high-dimensional parameter volumes, we use Monte Carlo Markov Chain (MCMC) techniques \citep{christensen01, lewis02b}.  Rather than computing new Markov chains, we importance-sample existing chains using the SPT lensing likelihood (e.g.,~Appendix B of \citealt{lewis02b}).  The chains we use were generated for the CMB temperature power spectrum analysis of K11 using a modified version of the CosmoMC package.  They provide full explorations of the allowed parameter volumes for various models, constrained by the WMAP7 CMB power spectrum measurements \citep{komatsu11}.  In some cases, we also consider the impact of including the SPT high-$\ell$ CMB temperature power spectrum measurements of K11.

The base parameter set varied in the chains consists of $(\Omega_bh^2, \Omega_ch^2, \theta_s,\tau,A_s, n_s)$, where $\theta_s $ is the angular scale subtended by the sound horizon at the CMB recombination surface.  The parameters describing the power spectrum of primordial fluctuations, $A_s$ and $n_s$, are defined relative to a reference wavenumber of $k = 0.002\,$Mpc$^{-1}$, as is chosen in the analysis of the WMAP team \citep{komatsu11}.  In the case of the chains which are computed with K11 data, the amplitudes of the three sources of foreground fluctuations which become important on small angular scales are also varied and marginalized over.  These consist of the amplitude of the power spectrum of clustered infrared galaxies; the amplitude of the power spectrum associated with the Poisson, or shot noise, nature of the galaxy distribution; and the amplitude of the power spectrum of  Sunyaev-Zel'dovich fluctuations.   All cosmological
parameters are assigned flat priors, with the exception of the 
logarithmic prior assigned to $A_s$. Foreground parameters have priors 
based on the measurements in \citet{shirokoff11}, as
described in K11.

We  generate a lensing power spectrum for each point in these chains.
The calculation of accurate theoretical lensing power spectra is CPU-intensive. For efficiency, we first calculate lensing power spectra for $\sim 10^4$ points in a chain using the CAMB software package \citep{lewis99}.  For this step we use the parameter chains provided by the  WMAP team as a training set.  
We then use this information to interpolate power spectra at other points in the parameter space. We perform a principal component analysis on the training set, keeping the first 12 modes. We then perform a linear fit to the amplitudes of these modes as a function of cosmological parameters. With this linear fit for the mode amplitudes we can construct a lensing power spectrum for any set of cosmological parameters. 
We find the fit (at $L<2000$) to have an rms difference of less than 1\% 
from the CAMB-computed power spectrum for cosmological models within the WMAP7-allowed $3\,\sigma$ parameter space.

For each of these theoretical lensing  power spectra, we then calculate the likelihood of the SPT lensing data.  We assume a Gaussian likelihood function, consisting of the SPT bandpowers shown in Figure~\ref{fig:cl_compilation} together with the covariance matrix used to constrain \alens\ above.  Unlike when fitting for the template amplitude $\alenszero$, in which we fit for each field separately, here we fit directly to the field-combined bandpowers.  

In the Metropolis-Hastings technique of MCMC integration \citep{metropolis53}, which is used for the chains considered in this paper, each location in parameter space examined by the chain is assigned a weight according to the number of iterations that the chain remained at that point.  To include the SPT lensing measurements, we calculate new weights by multiplying this weight by the SPT lensing likelihood.  We can then compute statistics, such as marginalized one- and two-dimensional parameter distributions, by replacing the original weights with these new weights.
The constraints on cosmological parameters presented in the remainder of this section are quoted as the means and variances of the distributions obtained using these new weights.  The constraints we obtain for each model are summarized in Table~\ref{tab:cosmo_results}.

\subsubsection{\LCDM}
\begin{figure}
  \plotone{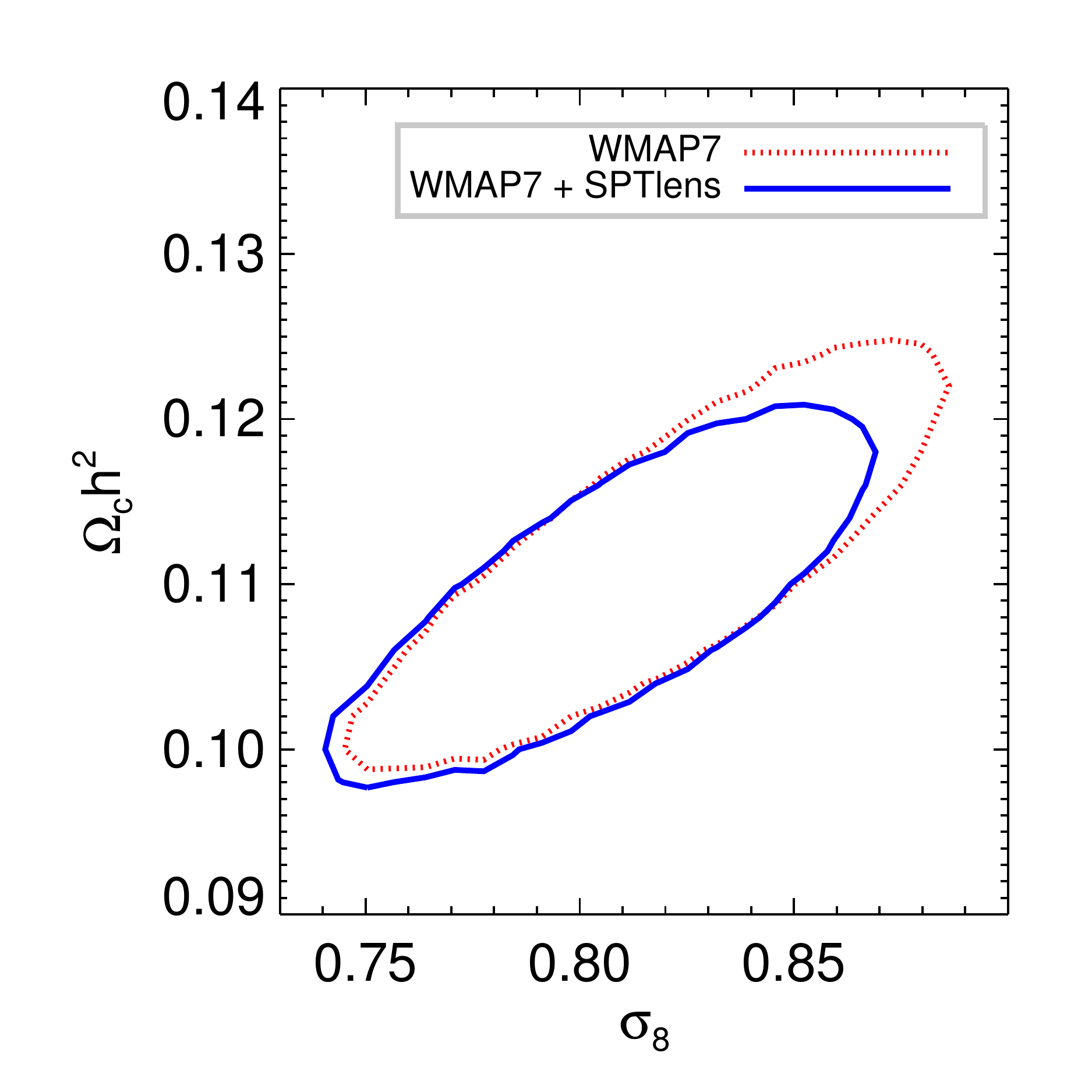}
  \caption{ 95\% confidence-level constraints on $\sigma_8$ and $\Omega_ch^2$ from WMAP7 data
    alone (red dotted contour), and improvement when including the SPT lensing data
    (blue solid contour). }
  \label{fig:onepanel_sigma8_omegac}
\end{figure}

The first cosmological model we consider is the spatially flat,
power-law LCDM model (with the lensing amplitude set to unity).  The parameter constraints which are most
improved when the SPT lensing data are added to the WMAP7-allowed \LCDM parameter volume are those of the cold
dark matter density and the two parameters related to the primordial
scalar fluctuation power spectrum.  In
Figure~\ref{fig:onepanel_sigma8_omegac}, we show associated
constraints on $\omegac$ and $\sate$, which is a derived quantity
given by the square root of the variance of the linearly-evolved
density field today in spheres of size 8$h^{-1}$\,Mpc.  The constraint
on $\sigma_8$ improves by $\sim 10\%$, from $\sigma_8 = 0.821 \pm
0.029$ to $\sigma_8 = 0.810 \pm .026$ compared with WMAP7 alone.

Zahn et al.~(2012, in preparation) suggest that, 
for the SPT data considered here, the lensing information contained in CMB power spectrum estimators and that contained in trispectrum estimators are largely independent.  
Combining the K11 temperature power spectrum data with the lensing bandpowers, 
we obtain a constraint on the matter fluctuation amplitude of $\sate = 0.814 \pm 0.020$, an improvement in precision of $\sim 30\%$ compared to WMAP7 alone.

\subsubsection{\alens}
\begin{figure}
  \centering
  \plotone{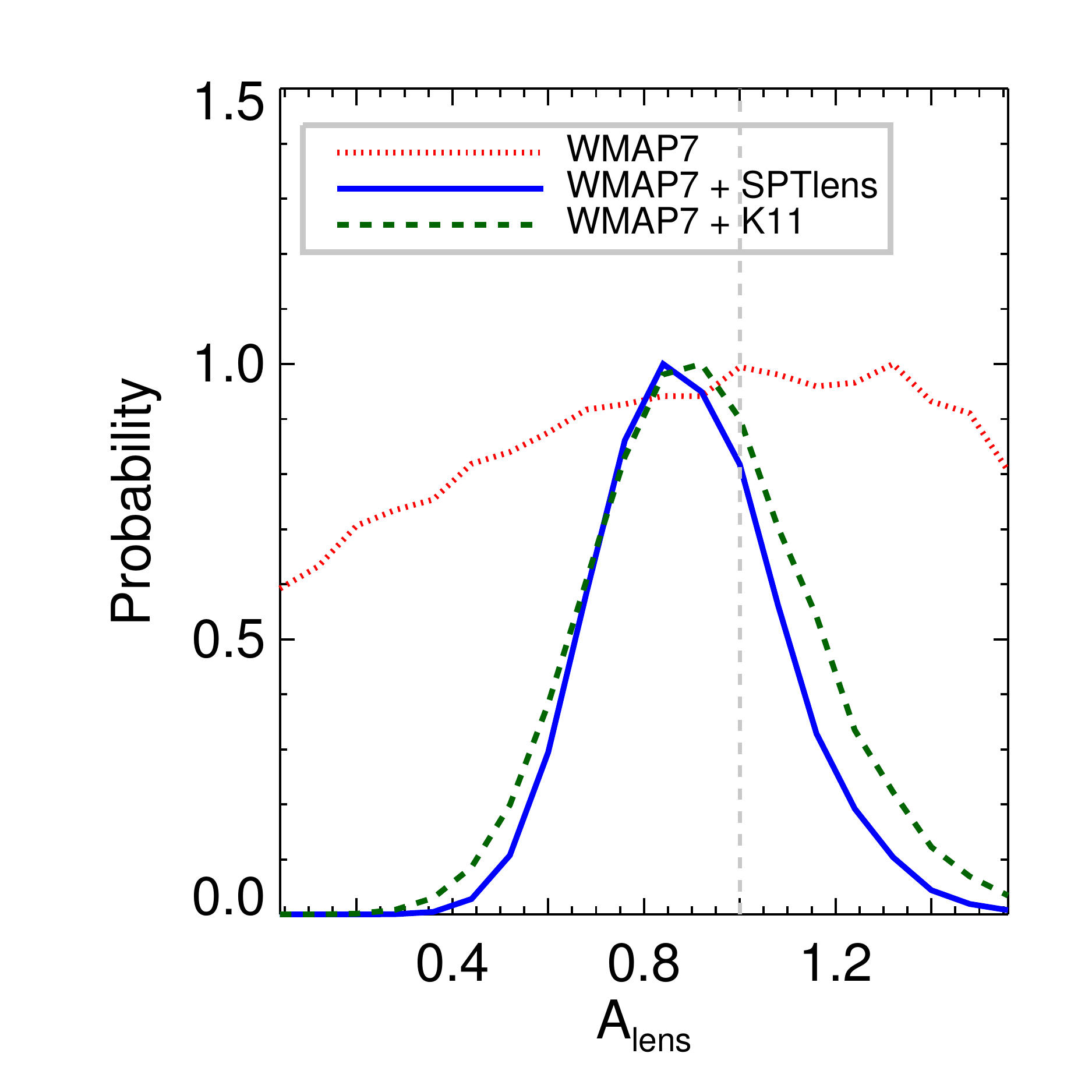}
  \begin{centering}

    \caption{One-dimensional constraints on the lensing amplitude when marginalizing over \LCDM parameters.  The WMAP power spectrum data (red dotted line) do not detect lensing.  The SPT lensing data show a clear detection of lensing (solid blue curve), as do the SPT temperature power spectrum data from K11 (green dashed line). }
    \label{fig:alens_mcmc}
  \end{centering}
\end{figure}

In Section~\ref{sec:lensamp} we performed a fit for the lensing amplitude at a fixed reference cosmology.  Here, we use MCMC techniques to find constraints on the lensing amplitude when marginalizing over \LCDM parameters.  At each point in the WMAP7 \LCDM chain, we define a parameter, $\alens$, which corresponds to the amplitude of the lensing  power spectrum relative to its value for the given set of \LCDM cosmological parameters.  We can then find constraints on this parameter, to which we assign a flat prior, jointly with the \LCDM parameters.  K11 used this approach to measure the lensing amplitude at high significance.  We find that the SPT lensing data in combination with WMAP7 measure the lensing amplitude  to be $\alens = 0.90 \pm 0.19$. 
The equivalent measure of the lensing impact on the temperature power spectrum from K11 is $\alens = 0.92 \pm 0.23$.   These constraints are shown in Figure~\ref{fig:alens_mcmc}.  
Combining the SPT lensing data with K11, neglecting any possible
correlation between the lensing information, gives $\alens = 0.90 \pm 0.15$, with  the six \LCDM parameters marginalized. 

\subsubsection{Curvature}
\label{sec:curvature}

Observations of the primary CMB at $z\sim 1100$ do not measure the
spatial curvature of the Universe to high precision.  This is due to
the angular diameter distance degeneracy.  A key physical length scale
associated with the observed CMB surface is the acoustic scale, and
observations of the CMB that include the acoustic peak region of the
power spectrum can measure the angular size corresponding to this
physical scale to high accuracy.  Indeed, the parameter $\theta_s$ is
used as one of the standard base parameters in cosmological fitting.
There is an effective degree of freedom associated with the angular
diameter distance to the CMB recombination surface, which is required
to convert the angular size to a physical length scale.  In the flat
\LCDM model (as we parameterize it), $\Omega_\Lambda$ plays this
role, and is well-constrained from primary CMB data (to $\sim 4\%$),
despite the fact that the dynamical effects of dark energy become
important long after last scattering.

\begin{figure}
  \plotone{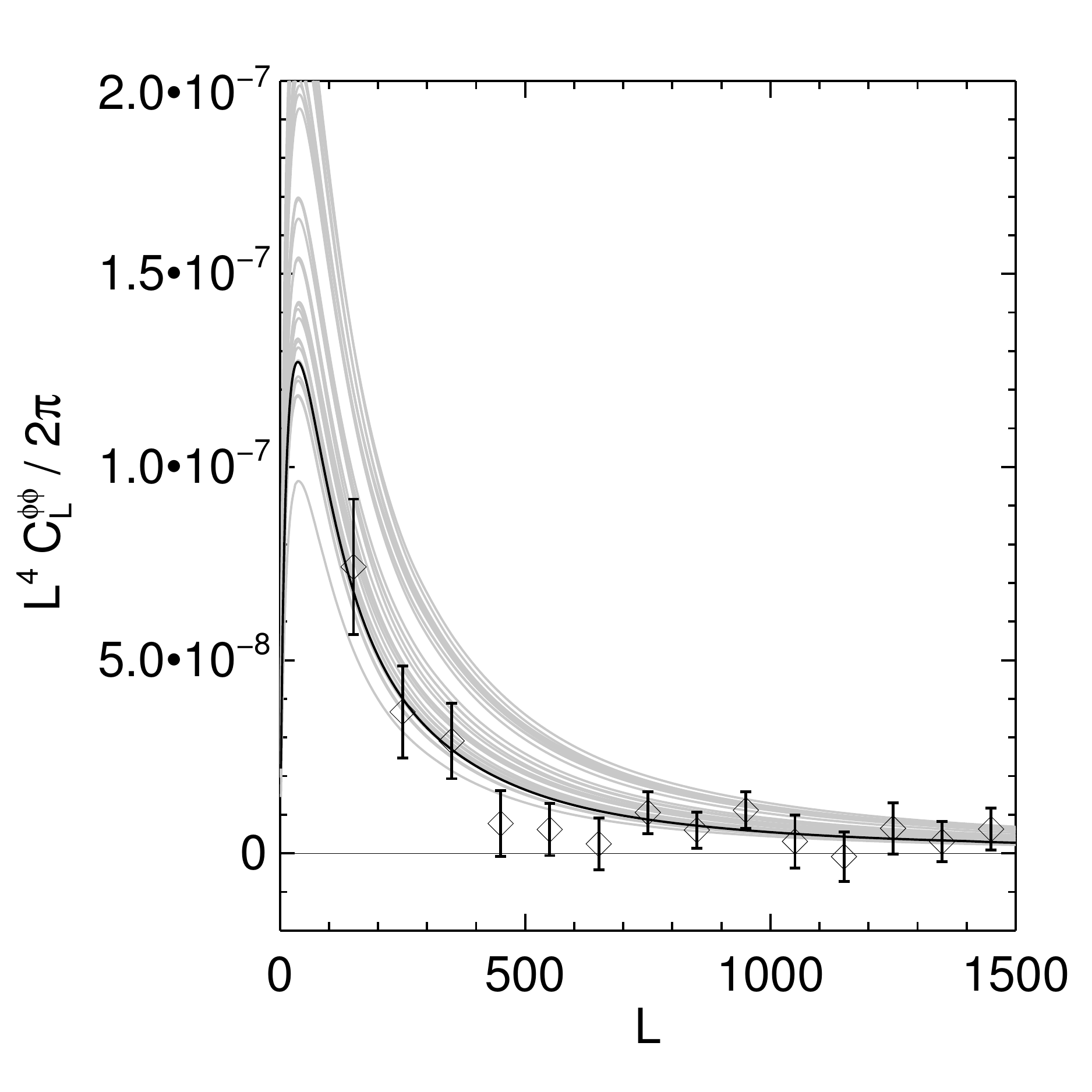}
  \begin{centering}
    \caption{Demonstrating the ability of the lensing data to constrain the free-curvature model.  Each grey line is the lensing  power spectrum for a point in cosmological parameter space allowed by WMAP7 when the curvature parameter is allowed to vary.  Specifically, the curves are taken from equally-spaced points in a WMAP7-only Markov chain which explores the \LCDM+$\Omega_k$ model.  The SPT lensing bandpowers, shown as black diamonds, can significantly discriminate among the models.  The lensing power spectrum for the fiducial cosmology, which has $\Omega_k = 0$, is shown as the black curve. }
    \label{fig:omegaksamples}
  \end{centering}
\end{figure}

However, in cosmological models which allow for an additional free
parameter that affects the angular size of the sound horizon, such as
curvature, only a particular linear combination of these parameters
will be well-constrained with primary CMB data, leading to a strong
parameter degeneracy.  Adding a measurement of the distance scale to
another redshift range, such as  that containing the matter fluctuations responsible for CMB
lensing, can break this degeneracy \citep[e.g.,][]{smith06} as was
shown experimentally by \citet{sherwin11}.  Indeed,
Figure~\ref{fig:omegaksamples} demonstrates that models with negative 
curvature, which are allowed by primary CMB temperature measurements,
can predict CMB lensing potential power spectra that are up to a
factor of two higher in amplitude than those predicted by \LCDM in
flat geometries.

\begin{figure*}
  \subfigure{\includegraphics[scale=\fourfigsscale]{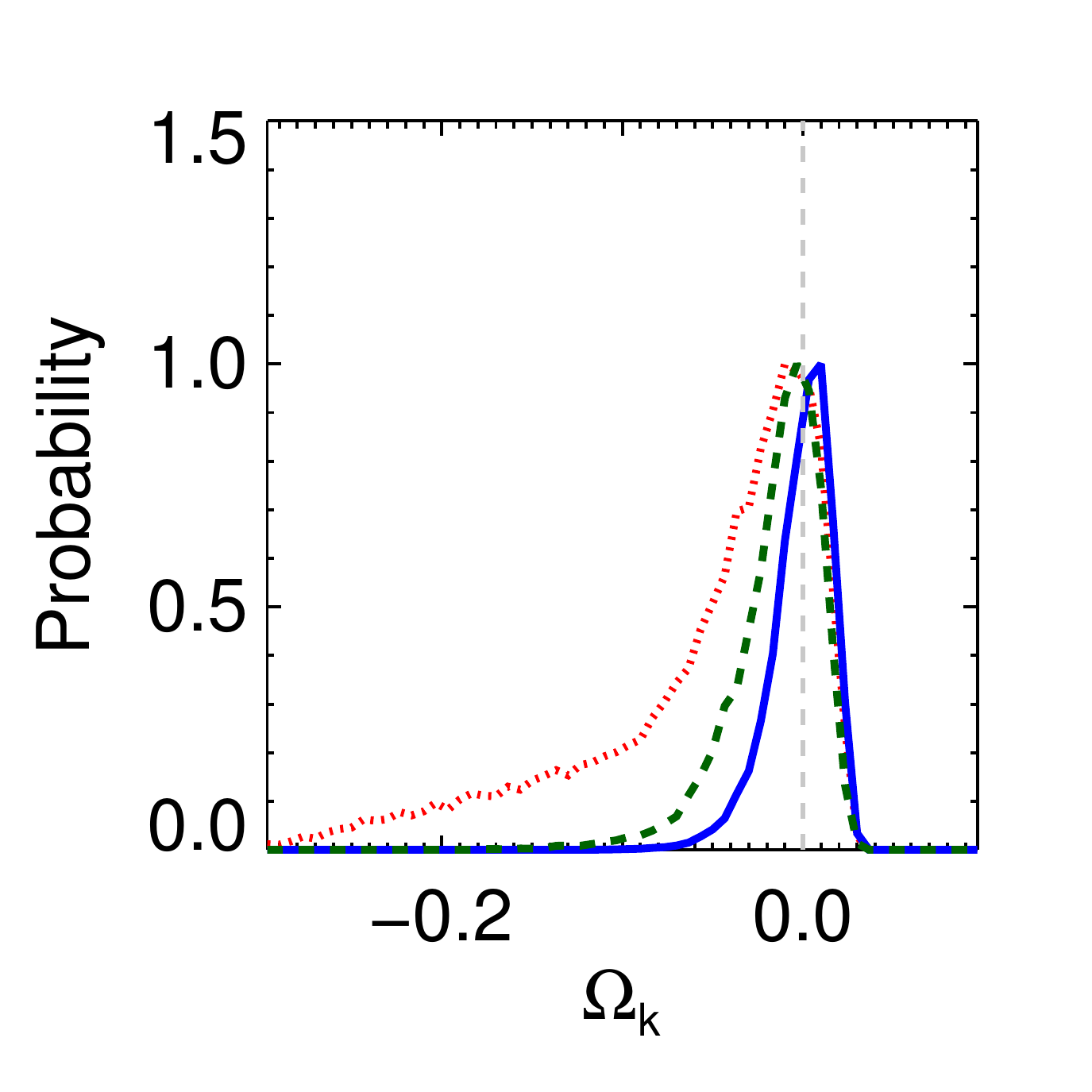}}
  \hspace{\fourfigshspace cm}
  \subfigure{\includegraphics[scale=\fourfigsscale]{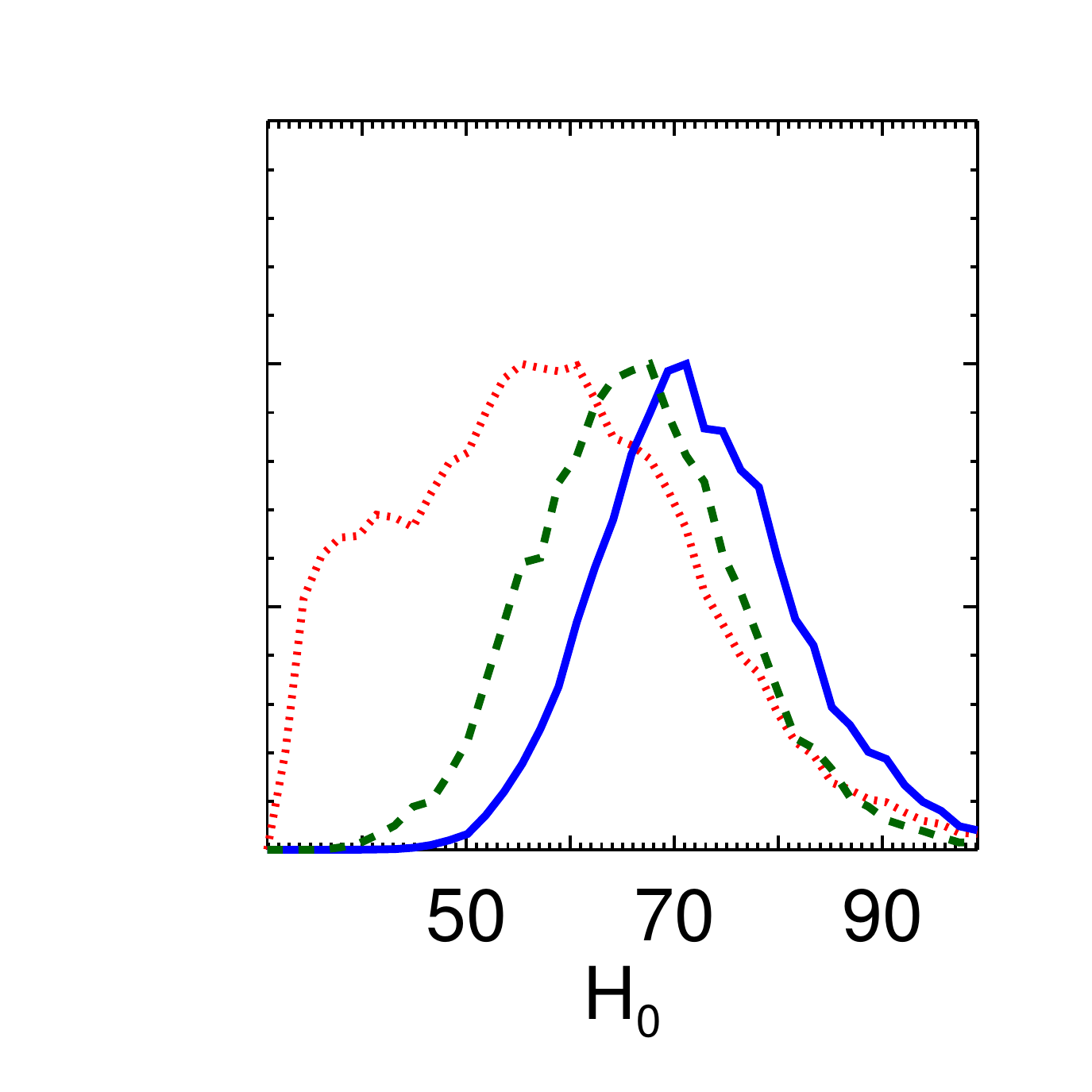}}
  \hspace{\fourfigshspace cm}
  \subfigure{\includegraphics[scale=\fourfigsscale]{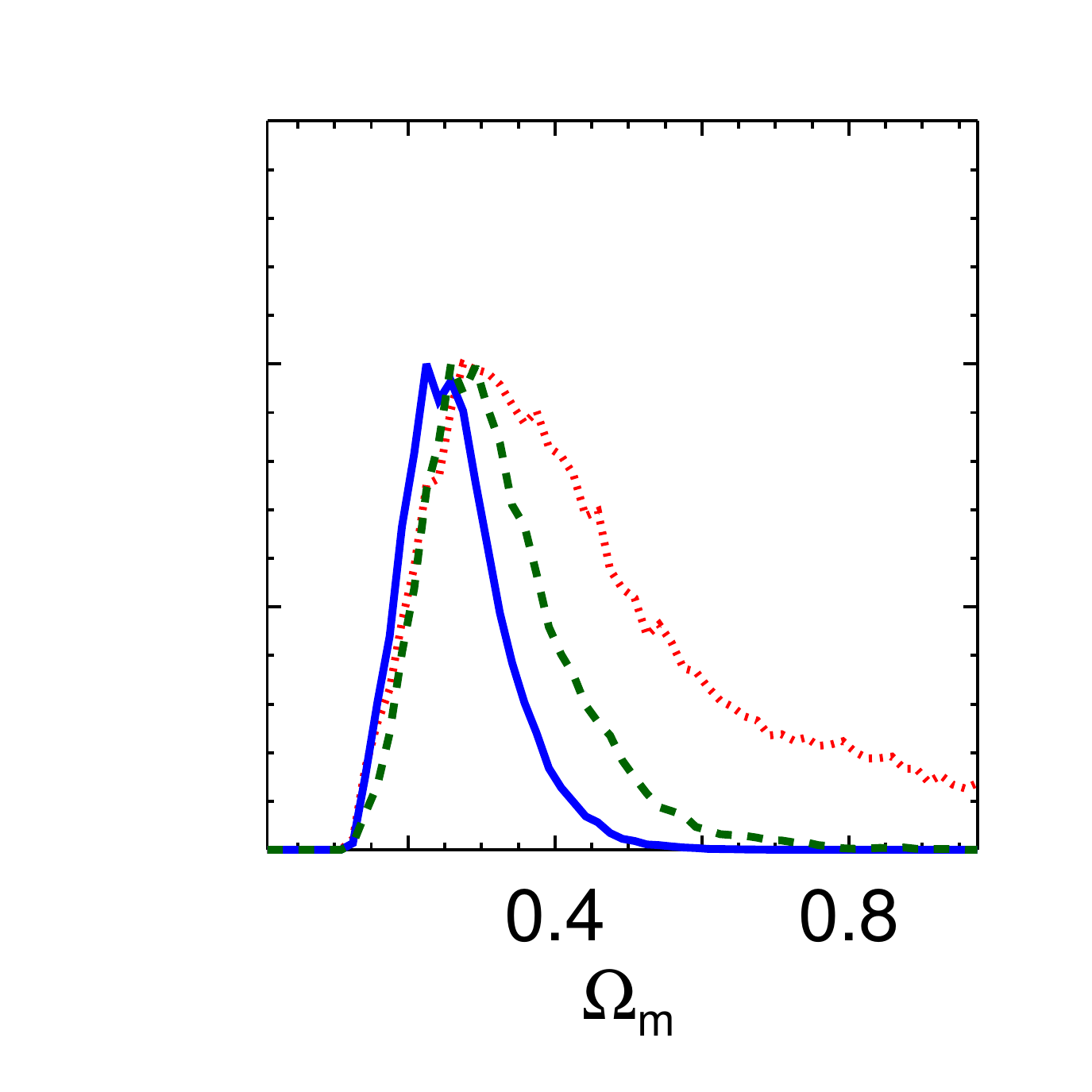}}
  \hspace{\fourfigshspace cm}
  \subfigure{\includegraphics[scale=\fourfigsscale]{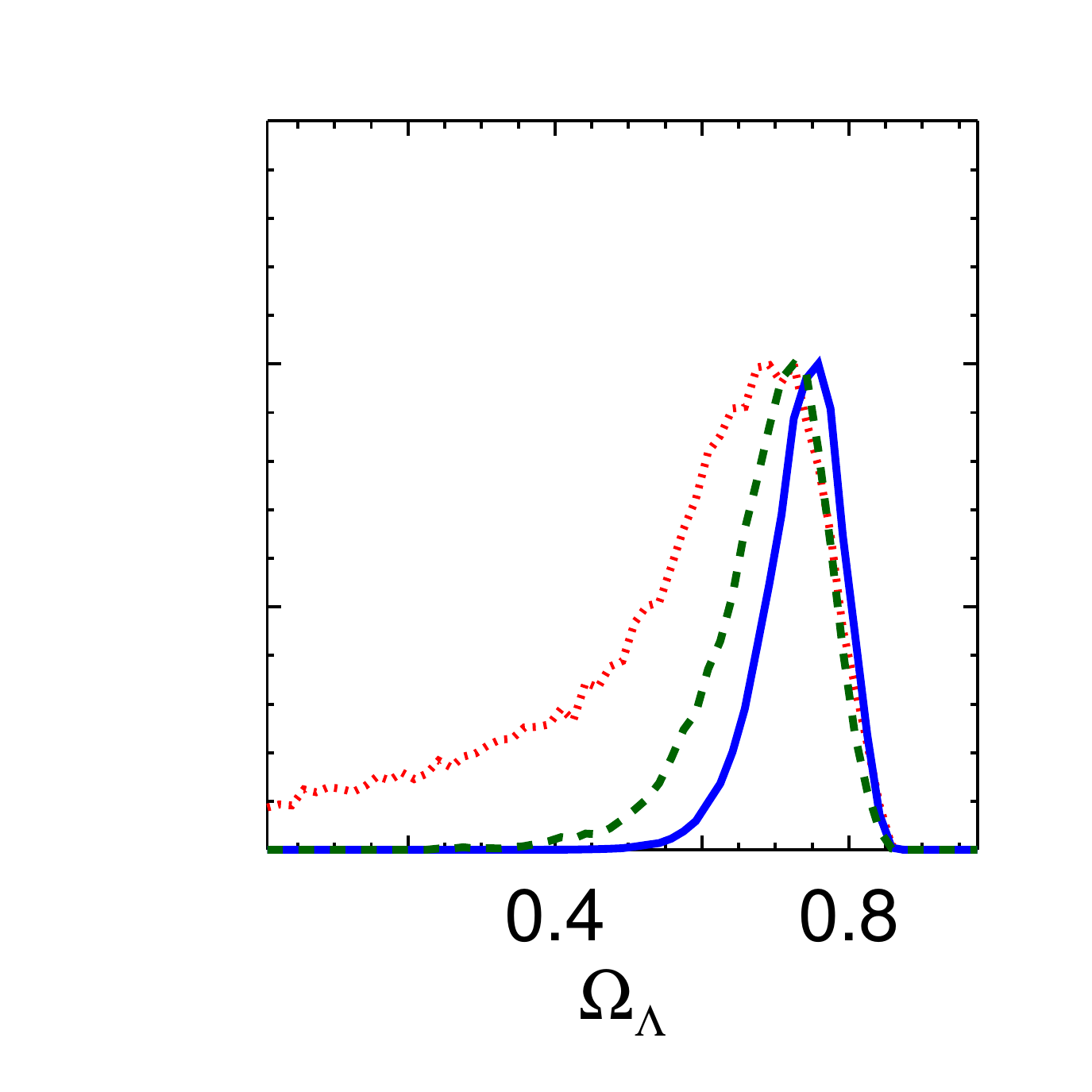}}
  \hspace{\fourfigshspace cm}
  \subfigure{\includegraphics[angle = 90, width=\textwidth]{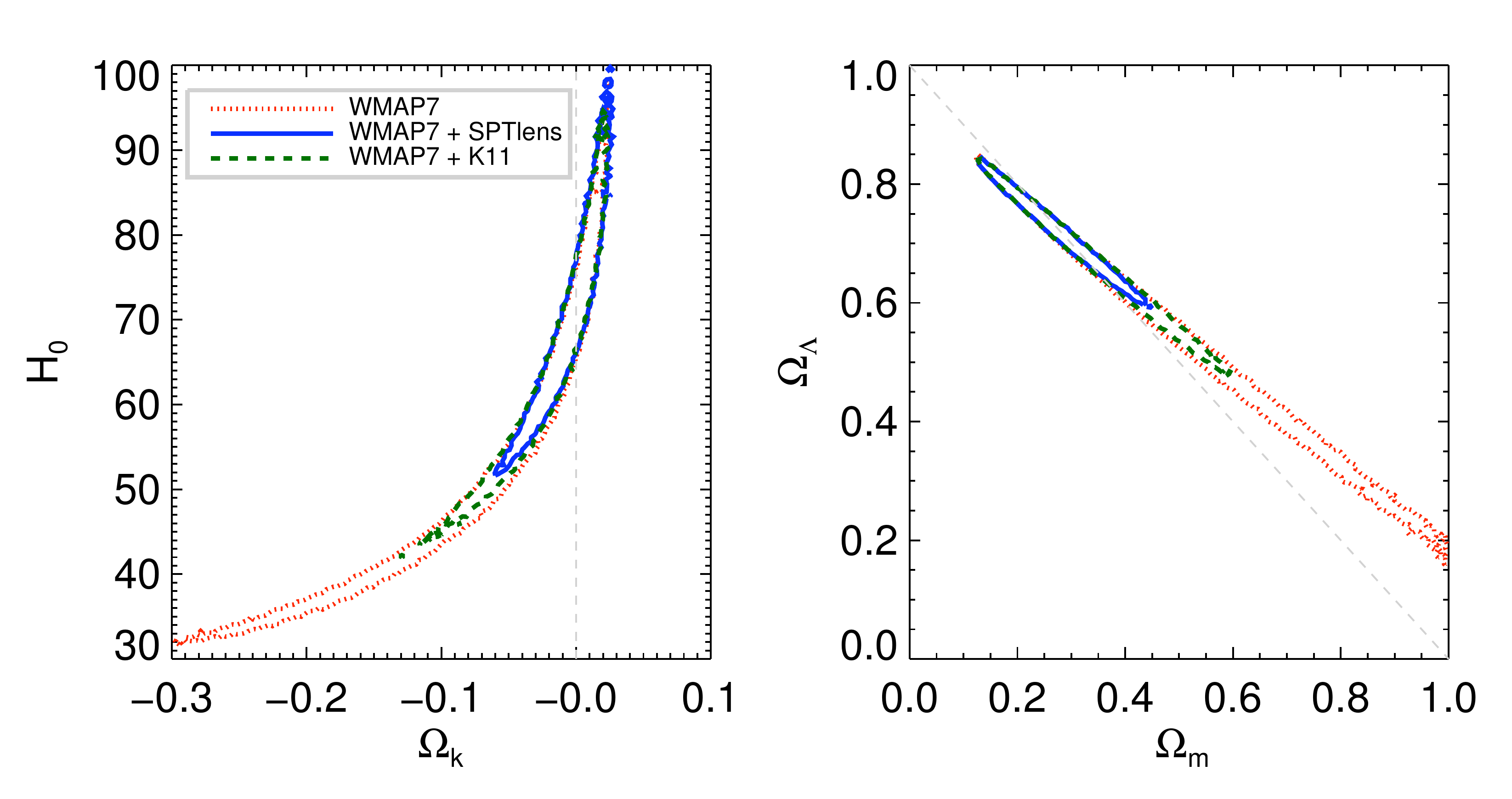}}
  \caption{Constraints on the free-curvature model \LCDMx+$\Omega_k$ from WMAP7 data alone (red dotted curves); adding the SPT lensing bandpowers to WMAP7 (blue solid curves); and adding the K11 temperature power spectrum measurements to WMAP7 (green dashed curves).  The top row of panels shows the one-dimensional parameter likelihoods on $\Omega_k$, $H_0$, $\Omega_m$, and $\Omega_\Lambda$; the two panels on the bottom  show the 95\% confidence-level contours in the $H_0$-$\Omega_k$ and $\Omega_\Lambda$-$\Omega_m$ planes.}
\label{fig:constraint_omegak_H0.pdf}
\end{figure*}

Here, we evaluate the improvement in the measure of the curvature
parameter $\Omega_k$ when the SPT lensing data are considered in
combination with the WMAP7 data, by adding the constraints from the
SPT lensing bandpowers to the WMAP7-allowed \LCDM+$\Omega_k$ parameter
volume.  We find the marginalized $1\,\sigma$ curvature constraint to
tighten by a factor of $\sim 3.9$ over WMAP7 alone, to $\Omega_k =
-0.001 \pm 0.017$.  Many of the models allowed by WMAP7 correspond to
values of the Hubble parameter $H_0$ as low as 30\,\hunit; adding the
SPT lensing data leads to an effective measure of $H_0 = (72.3 \pm
9.3)\,$\hunit, from the CMB alone.  This result is not currently competitive with
direct measures of the Hubble constant, which have an uncertainty of
2.4\,\hunit \citep{riess11}, but is of interest because the constraints
come only from the CMB.  These results are shown in
Figure~\ref{fig:constraint_omegak_H0.pdf}.  The constraint from the CMB
lensing measurement also corresponds to a measure of a nonzero dark
energy density, using only the CMB, of $\Omega_\Lambda = 0.734 \pm
0.056$.  Such a measure is not possible using only the primary CMB
anisotropies at recombination, without additional information from
lensing.  

The greater-than-$5\,\sigma$ constraint of the lensing amplitude  found using the lensing effect on the CMB power spectrum by K11 
(when accounting for the non-Gaussian probability distribution in
\alens) is also able to provide significant constraints on this parameter volume.  This corresponds to a measurement of the curvature of the Universe using only the CMB power spectrum, i.e.,~without performing the trispectrum-based lensing reconstruction that is the focus of this paper.  As is shown with the green dashed curves in Figure~\ref{fig:constraint_omegak_H0.pdf}, the K11 temperature bandpowers together with WMAP7 constrain the dark energy density to $\Omega_\Lambda = 0.689 \pm 0.081 $, the Hubble parameter to $H_0 = 66.4 \pm 9.8$\,\hunit, and the curvature parameter to $\Omega_k = -0.015 \pm 0.026$.  We have checked that these results are almost entirely due to the lensing effect on the K11 temperature power spectrum measurements; the constraints on these parameters degrade to close to their WMAP7-alone values when the $\alens$ parameter is marginalized.  Measures of the dark energy from the CMB alone are thus possible without performing lensing reconstruction, using only the effects of lensing on the CMB temperature power spectrum.

\subsubsection{Neutrino masses}

Massive neutrinos damp the matter power spectrum on scales which are smaller than their free-streaming scale at the redshift at which they become non-relativistic.  For the case of three neutrinos with degenerate masses, an increase in the sum of the neutrino masses of 0.1 eV leads to a decrease of 5\%  in the matter power spectrum on scales of $k \gtrsim 0.05 h\,$Mpc$^{-1}$.  The matter power spectrum suppression leads to a comparable level of suppression in the CMB lensing power spectrum at  $L \gtrsim 100$ \citep{kaplinghat03, lesgourgues06}, corresponding to the entire SPT signal band.    High-significance CMB lensing measurements hold the promise to measure the sum of neutrino masses at the $0.05\,$eV level  \citep[e.g.,][]{lesgourgues06,deputter09}, the minimum required for at least one species by oscillation experiments \citep{Adamson:2008zt}.

We generate lensing power spectra from the WMAP7-allowed \LCDMx+$\sumnu$ parameter space and again compute SPT lensing likelihoods for each model.  
Although the SPT lensing data have the statistical power to improve the constraint on the sum of neutrino masses by $\sim 20\%$, they also show a mild preference for low values of both $\sate$ and $\omegac$, as seen (in the case of \LCDMx) in Figure~\ref{fig:onepanel_sigma8_omegac}.  Both of these parameters are degenerate with the neutrino masses.  The mild preference for low values of $\sate$ corresponds to a mild preference for larger values of $\sumnu$.  The net result is that the WMAP7-based 95\% confidence level upper limit of the sum of neutrino masses actually increases slightly, from $\sumnu < 1.10\,$eV to $\sumnu < 1.17\,$eV.  

A significant fraction of the parameter space allowed by WMAP7 corresponds to values of the Hubble parameter which are inconsistent with recent observations.  With the measure of the Hubble parameter of \citep{riess11} included with WMAP7, adding the SPT lensing data changes the 95\% confidence level upper limit from  $\sumnu < 0.36\,$eV to  $\sumnu < 0.38\,$eV.

\subsubsection{Dark energy equation of state}
\begin{figure}
  \plotone{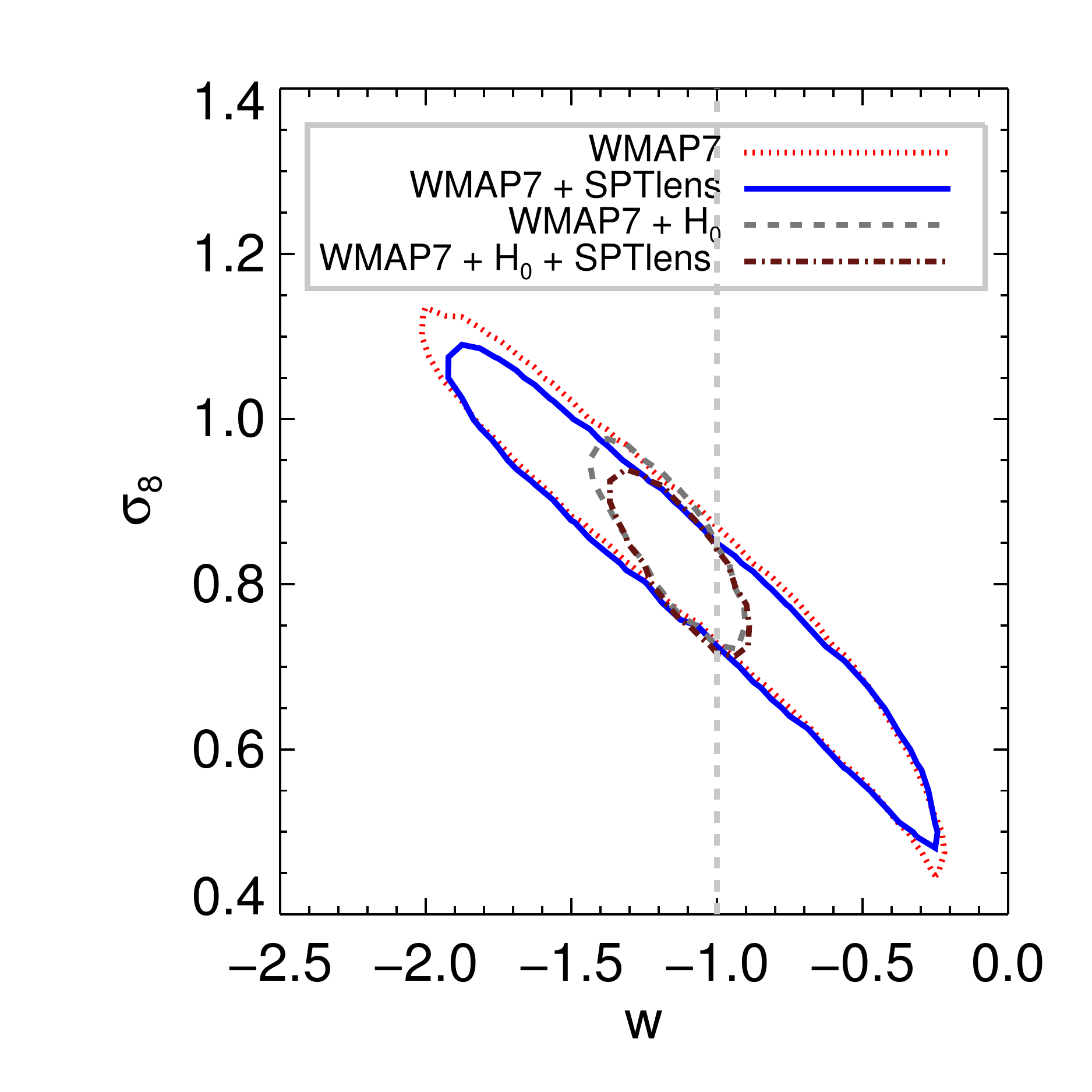}
  \begin{centering}
  \caption{95\% confidence level contours on the dark energy equation of state  $w$ and the matter fluctuation amplitude  $\sate$.  The WMAP7 data (red dotted contour) show a strong degeneracy between these two parameters.  Adding the SPT lensing data (blue solid contour) can start to break this degeneracy, tightening the $w$ constraint by 5\%.  When the measure of the Hubble parameter is also used (grey dashed contour), the lensing data improve the constraint by 15\% (brown dot-dashed contour).}
  \label{fig:onepanel_w_sigma8}
  \end{centering}
\end{figure}

The majority of the weight in the redshift kernel for CMB lensing, Eq.~\ref{eq:redshiftintegral}, lies in the matter-dominated era.  The amplitude of the lensing power spectrum can thus be used to provide a measure of the distance to these redshifts, leading to constraints on the equation of state of dark energy, $w$.  Assuming   $w$ to be constant as a function of redshift, we show the constraints in the $w$-$\sate$ plane in Figure~\ref{fig:onepanel_w_sigma8}.  WMAP7 weakly constrains $w$,  to $-1.120 \pm   0.420$, based on the measure of the expansion history provided by CMB observations.  Adding the SPT lensing data modestly improves this uncertainty on $w$, by 5\%.  

When including the  \citet{riess11} $H_0$ measurement together with WMAP7, the SPT lensing data improve the precision of $w$ by 15\%, from $w=  -1.126 \pm   0.111 $ to $w=  -1.087 \pm   0.096$.

\begin{table*}
\begin{center}
\begin{threeparttable}
\caption[Constraints on Cosmological Parameters using SPT Lensing Bandpowers]{Constraints on Cosmological Parameters using SPT Lensing Bandpowers}
\begin{scriptsize}  
\begin{tabular}{ | l  c ||  c  c | c  c | c c |}
\hline
Model and  & & WMAP7 & WMAP7     & WMAP7 + K11 & WMAP7 + K11          & WMAP7 + $H_0$ & WMAP7 + $H_0$   \\
 parameter & &       & + SPTlens &             & + SPTlens$  ^\dagger$ &               & + SPTlens\\
\hline \hline
\LCDM     &  \sate & $   0.821 \pm   0.029 $&$   0.810 \pm   0.026 $&  $   0.814 \pm   0.024 $&$   0.806 \pm   0.022 $&   $   0.809 \pm   0.027 $&$   0.803 \pm   0.025$  \\
(6-param.) & $\omc$ & $  0.1125 \pm  0.0054 $&$  0.1103 \pm  0.0047 $& $  0.1117 \pm  0.0048 $&$  0.1102 \pm  0.0042 $&$  0.1091 \pm  0.0043 $&$  0.1081 \pm  0.0039$  \\ 
\hline
\LCDMx+$\alens$ & $\alens$ & $    1.13 \pm    0.98 $&$    0.90 \pm    0.19 $&  $    0.92 \pm    0.23 $&$    0.90 \pm    0.15 $&$    1.26 \pm    1.03 $&$    0.95 \pm    0.19 $ \\
(7-param.) &&&&&&&\\
\hline
\LCDMx+$\Omega_k$ & $\Omega_k$ &$ -0.0545 \pm  0.0670 $&$ -0.0014 \pm  0.0172 $&$ -0.0150 \pm  0.0257 $&$ -0.0015 \pm  0.0146 $&$  0.0045 \pm  0.0053 $&$  0.0042 \pm  0.0052 $\\
(7-param.)& $H_0$ & $    57.6 \pm    13.8 $&$    72.3 \pm     9.3 $&$    66.4 \pm     9.8 $&$    72.2 \pm     7.9 $&$    73.3 \pm     2.4 $&$    73.6 \pm     2.4 $   \\
 & $\Omega_\Lambda $ &$   0.561 \pm   0.193 $&$   0.734 \pm   0.056 $&$   0.689 \pm   0.081 $&$   0.738 \pm   0.046 $&$   0.744 \pm   0.019 $&$   0.749 \pm   0.017 $ \\
\hline
\LCDMx+$\sumnu$ & \sumnu\ (eV) &     $< 1.10 $ (95\% CL) & $< 1.17 $ & $<1.34$& $<1.37$& $ < 0.36 $&$<   0.38 $\\
 (7-param.)& \sate & $   0.726 \pm   0.070 $&$   0.709 \pm   0.066 $&$   0.688 \pm   0.072 $&$   0.677 \pm   0.068 $&$   0.774 \pm   0.041 $&$   0.768 \pm   0.039$   \\
& $\omc$ & $  0.1187 \pm  0.0072 $&$  0.1184 \pm  0.0073 $&$  0.1208 \pm  0.0074 $&$  0.1212 \pm  0.0075 $&$  0.1094 \pm  0.0043 $&$  0.1088 \pm  0.0039$             \\
\hline
wCDM & $w$ & $  -1.120 \pm   0.420 $&$  -1.040 \pm   0.399 $&$  -1.160 \pm   0.363 $&$  -1.105 \pm   0.352 $&$  -1.126 \pm   0.111 $&$  -1.087 \pm   0.096$           \\
(7-param.)&  $\sate$ & $   0.854 \pm   0.143 $&$   0.818 \pm   0.131 $&$   0.863 \pm   0.120 $&$   0.838 \pm   0.115 $&$   0.863 \pm   0.053 $&$   0.838 \pm   0.045$   \\
& $\omc$ & $  0.1132 \pm  0.0056 $&$  0.1109 \pm  0.0048 $&$  0.1123 \pm  0.0047 $&$  0.1107 \pm  0.0042 $&$  0.1135 \pm  0.0056 $&$  0.1108 \pm  0.0047$              \\
\hline
\end{tabular}
\end{scriptsize}
\label{tab:cosmo_results}
\begin{tablenotes}
\item  Constraints on parameters of interest when SPT lensing information is added.  The three datasets to which we add SPT lensing constraints are the CMB power spectrum measurements from WMAP7 \citep{komatsu11}; WMAP7 together with the CMB power spectrum measurements from K11; and WMAP7 together with the measure of the Hubble parameter from \citet{riess11}. In the \LCDM case, 6 base parameters are varied and marginalized over in the MCMC; in the other models these 6 parameters are varied, plus either $\alens$, $\Omega_k$, $\sumnu$, or $w$.  All errors are $1\,\sigma$ standard deviations within Markov chains, weighted with the likelihoods for the given datasets.  The Hubble parameter is quoted in units of \hunit.  $^\dagger$The WMAP7+K11+SPTlens column is obtained by combining  the SPT CMB power spectrum measurements of K11 with the trispectrum-based lensing measure performed in this paper, and is subject to the validity of neglecting the covariance between the two measures (Zahn et al.~2012, in preparation).
\end{tablenotes}
\end{threeparttable}
\end{center}

\end{table*}

\section{Conclusions}
\label{sec:conclusions}

We have detected the power spectrum of gravitational lensing of the
CMB at high significance on scales of $8^\prime$ and larger $(L <
1500$).  We find  the amplitude of the measured signal in our
fiducial best-fit \LCDM cosmology to be $\alenszero = 0.86 \pm 0.16$.  This
detection represents an important step toward the eventual goal of
using the lensing of the microwave background as a precise probe of
the growth of structure and geometry of the Universe.

As part of this analysis, we
have modeled several important biases in lensing reconstruction, 
demonstrating the ability to remove the leading bias due to the Gaussian power in the map.
We have used two complementary approaches for dealing with this bias.  
In the first approach, we estimate the bias directly,
relying heavily on previous SPT results: the measured power spectrum
of the primary CMB (K11), backgrounds from dusty galaxies
and galaxy clusters \citep{lueker10, shirokoff11,reichardt11}, and the
known source counts in the maps \citep{vieira10}.  In principle, this
method leads to the maximum possible detection significance, as it
uses all of the available data.  However, since the Gaussian bias
exceeds the signal by a large factor over much of the lensing signal
band, uncertainties  due to the instrumental calibration and beam or small
uncertainties in power spectrum estimation lead to systematic uncertainty.  We
obtain a ${6.3}\,\sigma$ detection of CMB lensing while accounting
for these sources of systematic uncertainty.

In the second approach, lensing maps obtained from two disjoint
regions of Fourier space were cross-correlated. This ensured that
there was no Gaussian bias to remove \citep{hu01b, sherwin10}.  The
clear SPT detection of this signal provided a more direct indication of
lensing of the CMB.  This method is more robust to systematics, but
has less statistical significance (as implemented), providing a $3.9
\sigma$ detection.  However, the loss in signal-to-noise ratio is not
a fundamental property of the lensing measurement using this method.
More sophisticated techniques for dividing the Fourier domain into
several regions and combining the multiple quadratic pairings of these
regions should lead to an increased signal-to-noise ratio. Without a
large (and somewhat uncertain) noise bias to subtract, this is a
potentially cleaner signal for future measurements; experiments with
higher signal-to-noise ratio will require stricter
control of systematic uncertainties to subtract the noise bias to
substantially higher precision.

We have also extracted an estimate of the power spectrum of a
curl-like component in the lensing field.  This is a strong test of
our ability to measure the noise bias, since similar forms for the
noise bias appear in both the divergence and curl estimates.  By
detecting the curl-like component at the expected level, we have
passed a significant test of our understanding of the Gaussian
backgrounds and noise in the SPT experiment. Furthermore, we have
obtained $1.8 \sigma$ evidence of the lensing signal using this curl
estimator by itself: while the curl estimator is formulated to
reconstruct fields with the opposite parity than leading-order
lensing, higher-order lensing biases, similar to those found in
\citet{kesden03}, lead to a non-zero lensing signal in the curl
estimate.

The contamination of the lensing signal by non-Gaussianity in Galactic and extragalactic foregrounds was
simulated and found to be relatively small. The ability of SPT to
detect point sources down to relatively low flux levels has made it
possible to mask out point sources to a level where the point source
background becomes nearly Gaussian.  There is a residual bias
originating from the correlations between the point source field and
the lensing field; although in the current work it was neglected 
given our 15\% statistical uncertainty, this bias will need to
be better understood in future analyses.  In the case of SZ emission
from galaxy clusters, the lower-than-expected SZ signal measured by
\citet{lueker10,das11b,shirokoff11,dunkley11} and \citet{reichardt11}
means that only a handful of clusters need to be masked to reduce the
SZ contamination to a level that can be neglected for this analysis.

We have also investigated the constraints that our measurement of the
lensing power spectrum places on cosmological models.  We  found
that adding our measurement to those from WMAP7 improved the
precision of the measurement of the amplitude of matter density
fluctuations, $\sate$, by 10\%.  The lensing amplitude, marginalized
over WMAP7-allowed models, was found to be $\alens = 0.90 \pm 0.19$.
The lensing data are able to mildly break degeneracies in parameter
values that result from the analysis of primary CMB data, namely $w$
and $\sumnu$.  When also including external measures of the Hubble
parameter, the constraint on $w$ improved by 15\% when including the
SPT lensing data, to $w = -1.087 \pm 0.096$.  Additionally, as in
\citet{sherwin11}, we found that our measurement can break the angular
diameter distance degeneracy and constrain models with spatial
curvature. We found $\sigma(\Omega_k) = 0.017$ when combining with
WMAP7, and $\sigma(\Omega_k) \simeq 0.015$ when including the lensing
effect on the SPT CMB temperature power spectrum reported by K11.

Measurements of CMB lensing are expected to continue to rapidly
improve.  The recently-completed full SPT-SZ survey includes
approximately 2500 square degrees of CMB temperature measurements at
the same depth as those considered here, along with additional
measurements at 95 GHz and 220 GHz.  This survey should produce a
detection of the lensing signal at several times the significance of
the detection presented here.

The analysis of the full SPT-SZ survey will require a more careful
modeling of foreground astrophysics than we have performed here; for
this analysis we only included modes with $\ell <3000$ in the CMB maps
to avoid contamination by galaxies and galaxy clusters. However, there is
signal on smaller scales that can be recovered with a more careful
treatment of non-Gaussianity from foregrounds.  Upcoming polarization-sensitive CMB experiments (e.g.,\ SPTpol,
\citealt{mcmahon09}; ACTpol, \citealt{niemack10}; PolarBear,
\citealt{arnold09})
will reconstruct the lensing power spectrum with high signal-to-noise ratio,
but will need to deal with a distinct set of
systematic uncertainties \citep{su09, miller09}.
The robust detection presented here and the parameter constraints
that are enabled indicate that CMB lensing is 
emerging as a powerful probe of cosmology.

\acknowledgments

We thank E.~Anderes, S.~Das, O.~Dor\'e, D.~Hanson, W.~Hu, and B.~Sherwin 
for useful discussions.

The South Pole Telescope is supported by the National Science
Foundation through grants ANT-0638937 and ANT-0130612.  Partial
support is also provided by the NSF Physics Frontier Center grant
PHY-0114422 to the Kavli Institute of Cosmological Physics at the
University of Chicago, the Kavli Foundation and the Gordon and Betty
Moore Foundation.  The McGill group acknowledges funding from the
National Sciences and Engineering Research Council of Canada, Canada
Research Chairs program and, the Canadian Institute for Advanced
Research.  Oliver Zahn acknowledges support from an Inaugural Berkeley
Center for Cosmological Physics Fellowship.  R. Keisler acknowledges
support from NASA Hubble Fellowship grant HF-51275.01.  B.A. Benson is
supported by a KICP Fellowship.  M. Dobbs acknowledges support from an
Alfred P. Sloan Research Fellowship.  L. Shaw acknowledges the support
of Yale University and NSF grant AST-1009811.  M. Millea and L. Knox
acknowledge the support of NSF grant 0709498.  J. Mohr acknowledges
support from the Excellence Cluster Universe and the DFG research
program TR33 ``Dark Universe''. This research used resources of the
National Energy Research Scientific Computing Center, which is
supported by the Office of Science of the U.S. Department of Energy
under Contract No. DE-AC02-05CH11231.  It also used resources of the
CLUMEQ supercomputing consortium, part of the Compute Canada network.
We acknowledge the use of the Legacy Archive for Microwave Background
Data Analysis (LAMBDA).  Support for LAMBDA is provided by the NASA
Office of Space Science.  Some of the results in this paper have been
derived using the HEALPix \citep{gorski05} package.

\appendix

\section{Lensing Signature in the Curl Estimator Power Spectrum}
\label{sec:curlappendix}

\begin{figure}
  \centering
  \includegraphics[scale  = .6]{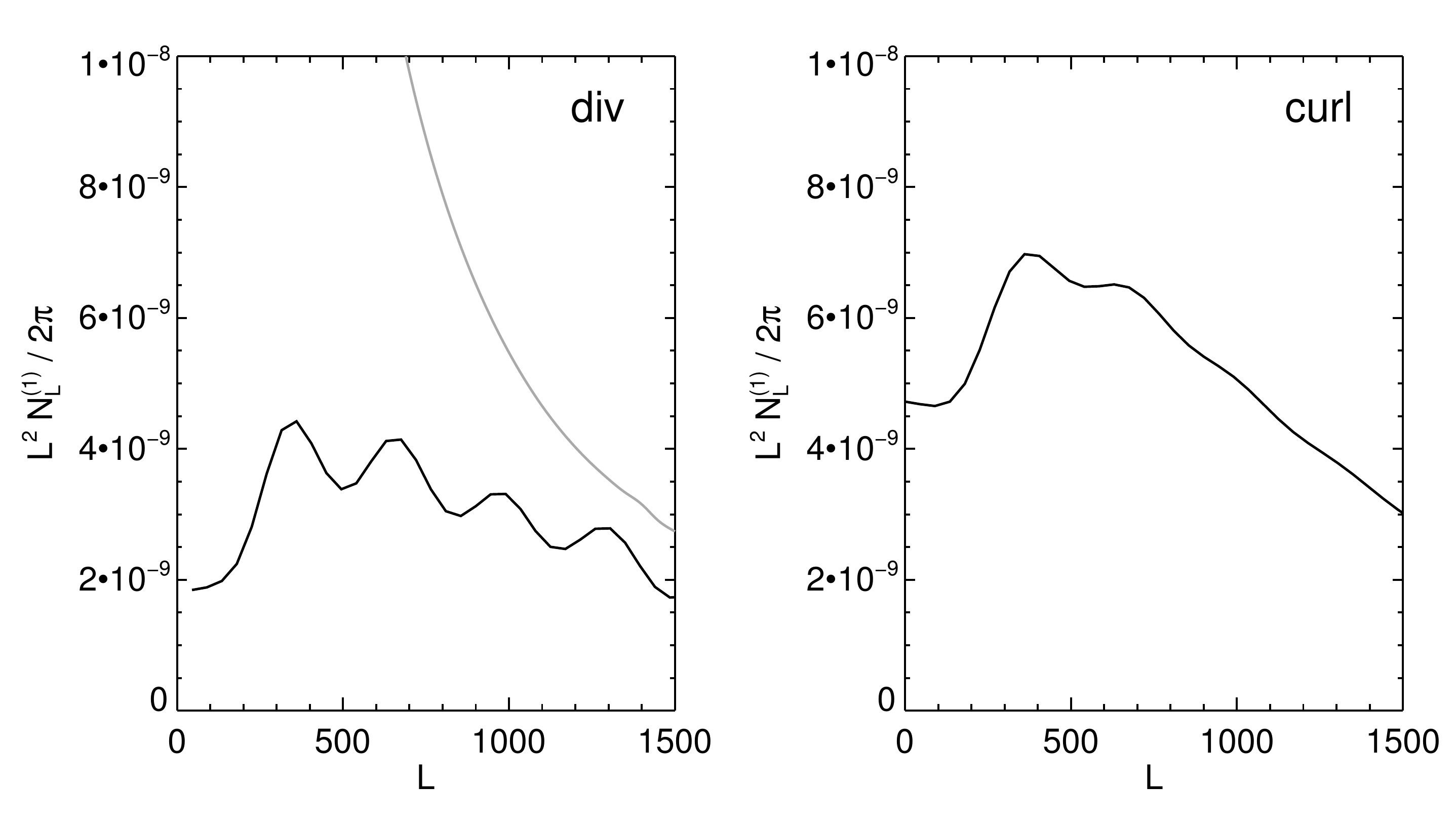}
  \caption{Higher-order bias signatures in the lensing estimate, described further in Section~\ref{sec:biases}.  Left: \citet{kesden03} bias in the divergence estimate (black) together with the lensing power spectrum (grey).  Right: similar bias in the curl estimate.  These are computed under the assumption of isotropic white noise and an analytical beam.}
\label{fig:ennoneplots}
\end{figure}

As discussed in Section~\ref{sec:curlestimator}, the curl estimator of
\citet{cooray05} is formulated to search for curl-like sources of
deflection in the CMB.  Instead of shifting the CMB by the gradient of
a scalar field $\phi$ according the usual lensing operation $T(\nhat) =
T^U(\nhat + \nabla \phi(\nhat))$, these sources, denoted $\Omega(\nhat)$, shift the CMB according to 
\begin{equation}
T(\nhat) = T^U(\nhat + \nabla \star \Omega(\nhat)).
\end{equation}
The operator $\star$ is given by $\mathbf{A} \star \mathbf{B} = A_yB_x
- A_xB_y$.  The signature of this mode of deflection is negligibly
small in a given reconstructed $\phi$ map.  However, additional terms
in the lensing trispectrum lead to the bias $\nlone$ in the estimated
power spectrum of this map, as they do for the divergence estimator of
\citet{kesden03}.  Indeed, in the main text we show that evidence for
this signal in the curl estimator is seen in the SPT data at
1.8 $\sigma$.  In Figure~\ref{fig:ennoneplots}, we
show the prediction, for both the divergence and curl components, under the assumptions of case of isotropic white noise and analytical beams.  In
practice, however, when presenting our results we compare against the
prediction for Monte Carlo estimates, in order to take into account
the anisotropic noise properties of the real dataset.

\bibliography{../../BIBTEX/spt}

\end{document}